\documentclass[useAMS,usenatbib]{mn2e}

\usepackage{epsfig,color,pifont,array}
\usepackage{wasysym}

\DeclareMathAlphabet{\mathpzc}{OT1}{pzc}{m}{it}

\usepackage{natbib}
\bibpunct{(}{)}{;}{a}{}{,}
\def\d{{\rm{d}}}
\def\ts{{\thinspace}}
\def\lb{{\langle}}
\def\rb{{\rangle}}
\def\llb{\left\langle}
\def\rrb{\right\rangle}

\def\spose#1{\hbox to 0pt{#1\hss}}
\def\ltsimm{\mathrel{\spose{\lower 3pt\hbox{$\sim$}}
        \raise 2.0pt\hbox{$<$}}}
\def\gtsimm{\mathrel{\spose{\lower 3pt\hbox{$\sim$}}
        \raise 2.0pt\hbox{$>$}}}

\def\cm{{\rm\thinspace cm}}

\def\s{{\rm\thinspace s}}

\def\g{{\rm\thinspace g}}

\def\erg{{\rm\thinspace erg}}
\def\Hz{{\rm\thinspace Hz}}

\def\ster{{\rm\thinspace ster}}
\def\ergps{\hbox{${\rm\erg\s^{-1}\,}$}}

\def\pcm{\hbox{${\rm\cm^{-1}\,}$}}
\def\pcm2{\hbox{${\rm\cm^{-2}\,}$}}
\def\pcm3{\hbox{${\rm\cm^{-3}\,}$}}
\def\ergpscm3Hz{\hbox{${\rm\ergps\cm^{-3}\Hz^{-1}\,}$}}
\def\ergpscm3Hzster{\hbox{${\rm\ergps\cm^{-3}\Hz^{-1}\ster^{-1}\,}$}}
\def\gpcm3{\hbox{${\rm\g\cm^{-3}\,}$}}
\def\ergpcm2{\hbox{${\rm\erg\cm^{-2}\,}$}}
\def\ergpcm3{\hbox{${\rm\erg\cm^{-3}\,}$}}
\def\phpscm2{\hbox{${\rm photons\s^{-1}\cm^{-2}\,}$}}

\def\aap{{\rm A\&A}}
\def\apj{{\rm ApJ}}
\def\apjl{{\rm ApJL}}
\def\apjs{{\rm ApJS}}

\def\mnras{{\rm MNRAS}}

\def\physrep{{\rm PhR}}

\def\jcp{{\rm J.~Comput.~Phys}}

\title [Global MHD disks II]{Global simulations of magnetorotational
  turbulence II: turbulent energetics}

\author[E.~R.~Parkin]
{E.~R.~Parkin\thanks{E-mail: ross.parkin@anu.edu.au} 
  \\Research School of Astronomy and Astrophysics,
  Australian National University, Canberra, ACT 2611, Australia}

\begin{document}

\date{Accepted ... Received ...; in original form ...}

\pagerange{\pageref{firstpage}--\pageref{lastpage}} \pubyear{2013}

\maketitle

\label{firstpage}

\begin{abstract}
  {Magnetorotational turbulence draws its energy from gravity and
    ultimately releases it via dissipation. However, the quantitative
    details of this energy flow have not been assessed for global disk
    models. In this work we examine the energetics of a well-resolved,
    three-dimensional, global magnetohydrodynamic accretion disk
    simulation by evaluating statistically-averaged mean-field
    equations for magnetic, kinetic, and internal energy using
    simulation data. The results reveal that turbulent magnetic
    (kinetic) energy is primarily injected by the correlation between
    Maxwell (Reynolds) stresses and shear in the (almost Keplerian)
    mean flow, and removed by dissipation. This finding differs from
    previous work using local (shearing-box) models, which indicated
    that turbulent kinetic energy was primarily sourced from the
    magnetic energy reservoir. Lorentz forces provide the bridge
    between the magnetic and kinetic energy reservoirs, converting
    $\sim 1/5$ of the total turbulent magnetic power input into
    turbulent kinetic energy. The turbulent energies (both magnetic
    and kinetic) are mainly driven by terms associated with the
    turbulent fields, with only a minor influence from mean magnetic
    fields. The interaction between mean and turbulent fields is most
    evident in the induction equation, with the mean radial magnetic
    field being strongly influenced by the turbulent electromotive
    force (EMF).

    During the quasi-steady turbulent state roughly $2/3$ of the
    Poynting flux travels into the corona, with the remainder
    transporting magnetic energy in the radial direction. In contrast
    to previous studies, the stress-related part of the Poynting flux
    is found to dominate, which may have important implications for
    ``reflection'' models of Seyfert galaxy coronae that typically
    invoke a picture of buoyant rising of magnetic flux tubes via
    advection.}
\end{abstract}




\section{Introduction}
\label{sec:intro}

Accretion disks possessing a weak initial seed magnetic field, a
radially decreasing angular velocity, and electrical conductivity are
unstable to the magnetorotational instability \citep[MRI
-][]{BH91,BH98}. Astrophysical disks, which typically have a Keplerian
rotation profile, satisfy this requirement. Following the linear phase
of instability growth, fully developed, self-sustaining turbulence
establishes \citep{Hawley:1995, Brandenburg:1995}. The same
  turbulent stresses that provide energy injection simultaneously
  drive angular momentum transport, thus enabling the maintenance of
  the turbulence in tandem with ongoing accretion \citep{BH98}.

Over the past two decades magnetorotational turbulence has taken
  central stage as the much sought angular momentum transport
  mechanism in accretion disks. However, transporting angular momentum
  is not the entire story. The very fact that the turbulence is
  self-sustaining raises questions about energetics. Primarily, what
  is the energy source, how (if at all) is energy removed from the
  disk, and what role does turbulence play? These questions have been
  broached previously but only using local (shearing-box) models of
  accretion disks \citep{Brandenburg:1995, Gardiner:2005b,
    Fromang:2007a, Simon:2009}. Recent work by \cite{Parkin:2013b} has
  highlighted important differences between local and global accretion
  disk models in terms of the influence of boundary conditions on both
  the mean field and magnetic energy evolution. This raises the
  question of whether or not the shearing-box approximation captures
  the physics of energy generation accurately. The establishment of an
  outer scale to the turbulence (i.e. a turn-over in the power spectra
  at low wavenumber) in the shearing-box simulations presented by
  \cite{Davis:2010} would suggest that the situation should not change
  considerably when moving to a global disk. Nevertheless, this
  intuition requires confirmation. With this in mind, in this work we
  examine the transport of energy using the results of a high
  resolution global accretion disk simulation coupled to a Reynolds
  averaged mean field analysis \citep[e.g.][]{BH98, Kuncic:2004}.

  In common with previous work we find that shear-stress correlations
  are at the heart of the turbulent energy source, both for the
  magnetic and kinetic energies \citep{Brandenburg:1995}. However, in
  contrast, we find that the turbulent kinetic energy gains the
  majority of its power from the hydrodynamic interaction between
  Reynolds stresses and mean flow shear, and is not primarily driven
  by Lorentz forces. Mean magnetic fields are not observed to directly
  influence the turbulent energies. However, a turbulent field is by
  definition a deviation from a mean field. The prominence of
  turbulent fields in driving mean field induction therefore
  highlights the mean-turbulent field interaction, which evidences the
  indirect nature by which mean magnetic fields influence turbulent
  energy evolution.

The remainder of this paper is organised as follows: in
\S~\ref{sec:model} we describe the simulation setup and averaging
procedures used in this investigation. In \S~\ref{sec:character} we
present basic characteristics of the mean and turbulent fields in the
disk. The results from the application of a control-volume analysis to
the simulations are presented in \S~\ref{sec:control_volume}. We
discuss the global energy flow and make a comparison with previous
work in \S~\ref{sec:discussion}, closing with conclusions in
\S~\ref{sec:conclusions}.

\section{The model}
\label{sec:model}

\subsection{Simulation code}
\label{subsec:hydromodel}

The time-dependent equations of ideal magnetohydrodynamics are solved
using the {\sevensize PLUTO} code \citep{Mignone:2007} in a 3D spherical
$(r,\theta,\phi)$ coordinate system. The relevant equations for mass,
momentum, energy conservation, and magnetic field induction are:
\begin{eqnarray}
\frac{\partial\rho}{\partial t} + \nabla \cdot (\rho {\bf v}) &  =  &
\dot{M}_{\rm source}, \label{eqn:mass_cons}\\
\frac{\partial\rho{\bf v}}{\partial t} + \nabla\cdot(\rho{\bf vv} -
{\bf BB}) + \nabla p & = & -\rho \nabla\Phi,\label{eqn:mom}\\
\frac{\partial E}{\partial t} + \nabla\cdot((E + p_{\rm tot}){\bf v} - ({\bf v \cdot B}){\bf B}) &
=& -\rho {\bf v}\cdot \nabla\Phi - \rho\Lambda \label{eqn:energy}\\
\frac{\partial\bf B}{\partial t} & = & \nabla \times ({\bf v \times B}). \label{eqn:induction}
\end{eqnarray}
\noindent Here $E = u_{\epsilon} + u_{K} + u_{\rm B}$ is the total
energy, $u_{\epsilon}=\rho \epsilon$ is the internal energy,
$\epsilon$ is the specific internal energy,
$u_{K}=\frac{1}{2}\rho|{\bf v}|^{2}$ is the total kinetic energy,
${\bf v}$ is the velocity, $\rho$ is the mass density, $p$ is the gas
pressure, $u_{\rm B}=\frac{1}{2}|B|^2$ is the magnetic
energy/pressure, and $p_{\rm tot} = p + u_{\rm B}$ is the total (gas
plus magnetic) pressure. We use an ideal gas equation of state, $p =
(\gamma - 1) u_{\epsilon}$, with an adiabatic index $\gamma =
5/3$. The adopted scalings for density, velocity, temperature, and
length are, respectively,
\begin{eqnarray}
  \rho_{\rm scale}&=&1.67\times10^{-7}~{\rm gm~s^{-1}}, \nonumber \\
  v_{\rm scale}&=&c, \nonumber \\
  T_{\rm scale}&=&\mu m c^{2}/k_{\rm B} = 6.5\times10^{12}~{\rm K},
  \nonumber \\
  l_{\rm scale}&=&1.48\times10^{13}~{\rm cm}, \nonumber 
\end{eqnarray}
\noindent where $c$ is the speed of light, and the value of $l_{\rm
  scale}$ corresponds to the gravitational radius of a $10^{8}~{\rm
  M_{\odot}}$ black hole.

The gravitational potential due to a central point mass situated at
the origin, $\Phi$, is modelled using a pseudo-Newtonian potential
\citep{Paczynsky:1980}:
\begin{equation}
\Phi = \frac{-1}{r - 2}.
\end{equation}
\noindent
Note that we take the gravitational radius (in scaled units), $r_{\rm
  g} = 1$. The Schwarzschild radius, $r_{\rm s}=2$ for a spherical
black hole and the innermost stable circular orbit (ISCO) lies at
$r=6$. 

A mass source term, $\dot{M}_{\rm source}$ is included in
Eq~(\ref{eqn:mass_cons}) which relaxes the gas density within a narrow
annulus spanning the outer $\sim12\%$ of the radial domain ($31 \leq r
\leq 34$, $|z|<2H$) towards the initial density distribution (see
\S~\ref{subsec:initialconditions}) over a timescale of an orbital
period, where $H$ is the thermal disk scale-height. Including
$\dot{M}_{\rm source}$ allows the total disk mass to reach a
quasi-steady value, thus enabling long simulation runs. The $\Lambda$
term on the RHS of Eq~(\ref{eqn:energy}) is an ad-hoc cooling term
used to keep the scale-height of the disk approximately constant
throughout the simulations by driving the temperature distribution in
the disk back towards the initial one over a timescale of an orbital
period; without any explicit cooling in conjunction with an adiabatic
equation of state, dissipation of magnetic and kinetic energy leads to
an increase in gas pressure and, consequently, disk scale-height over
time. The form of $\Lambda$ is particularly simple
\citep{Parkin:2013},
\begin{equation}
  \Lambda = \frac{1}{(\gamma - 1)} \frac{T(R,z) -
    T_{\rm 0}(R)}{2 \pi R/v_{\phi}} \label{eqn:cooling_function}
\end{equation}
\noindent where $T_{0}(R)$ and $T(R,z)$ are the position dependent
initial and current temperature, respectively, $v_{\phi}$ is the
rotational velocity, and $R$ is the cylindrical radius. Radiative
cooling is applied using an operator-split approach at second-order
accuracy.

The {\sevensize PLUTO} code was configured to use the five-wave HLLD
Riemann solver of \cite{Miyoshi:2005}, piece-wise parabolic
reconstruction \citep[PPM -][]{Colella:1984}, limiting during
reconstruction on characteristic variables \citep[e.g.][]{Rider:2007},
second-order Runge-Kutta time-stepping, the upwind Constrained
Transport scheme \citep{Gardiner:2005}, and the FARGO-MHD module
\citep[which permits larger time steps in problems involving rapid
rotation - see][for further details]{Mignone:2012}. For the FARGO-MHD
module the background rotation profile was fixed to Keplerian
rotation. To aid code stability, a small artificial viscous flux
\citep[with coefficient 0.1 - see][]{Colella:1984}, which only
switches on in regions of strong compression, is added to the Riemann
solver fluxes. In addition to the above, to reduce the impact of high
Alfv{\' e}n speeds in the low density coronal region we have modified
the {\sevensize PLUTO} code to include the approximate Alfv{\' e}n
speed limiter suggested by \cite{Gombosi:2002}. This latter
modification equates to replacing the conservative finite volume time
update of Eq~(\ref{eqn:mom}) with:
\begin{equation}
  \left(\left(1+ \frac{v_{\rm A}^2}{c_{\rm lim}^2}\right)\rho {\bf
      v}\right)_{n+1}= \left(\left(1+ \frac{v_{\rm A}^2}{c_{\rm
        lim}^2}\right)\rho {\bf v}\right)_{n} + \Delta t \times ...
\end{equation}
where $v_{\rm A}$ is the Alfv{\' e}n speed, $c_{\rm lim}$ is the
reduced speed of light, and the subscript $n$ denotes the time step
number. The above limiter has the advantage that steady-state
solutions are independent of the choice of $c_{\rm
  lim}$. Nevertheless, we choose the conservative value $c_{\rm
  lim}=0.1$, which is sufficient to aid with high values of $v_{\rm
  A}$ whilst being larger than the other wave speeds in the
simulation.

The simulation grid uses ($n_{r},n_{\theta},n_{\phi}$) = (512,\ts
256,\ts 256) uniformly spaced cells, covering the spatial extent
$8<r<34$, $\pi/2 - \theta_{0} <\theta< \pi/2 + \theta_{0}$ (where
$\theta_{0}=\tan^{-1}(3 H/R )$), and $0< \phi < \pi /2$. (The
$\theta$-extent of the grid equates to $|z|\pm 3H$ for a constant
aspect ratio disk.) In terms of cells per scale height, the adopted
grid has ($n_{r}/H,n_{\theta}/H,n_{\phi}/H) \simeq (16-67, 43, 16)$
which is within the regime of convergence as a function of resolution
found by \cite{Parkin:2013b}. The adopted boundary conditions are
identical to those used in \cite{Parkin:2013} with the exception that
for the $\theta$-boundaries we reflect density, pressure, and the
normal velocity if the cell adjacent to the boundary is
inflowing. This latter modification allows outflow whilst reducing
spurious fluctuations in the cells adjacent to the $\theta$-boundary
\citep{Flock:2011}. A buffer zone, similar to that used by
\cite{Fromang:2006} and \cite{Flock:2011} is used between $8 \leq r
\leq 12$ that has logarithmically increasing resistivity ($\eta$) and
viscosity ($\nu$) with $\eta_{\rm min}=\nu_{\rm min}= 10^{-6}$ and
$\eta_{\rm max} = \nu_{\rm max} = 10^{-4}$. The resistive term is
integrated explicitly, whereas super time stepping \citep[see][and
references therein]{Mignone:2007} is used to integrate the viscous
term. Finally, floor density and pressure values are used which scale
linearly with radius and have values at the outer edge of the grid of
$10^{-3}$ and $1.5\times10^{-6}$, respectively.

With the combination of changes to the simulation algorithms described
above \citep[relative to that used in][]{Parkin:2013, Parkin:2013b},
an increase in average time step size by a factor of roughly 8-10 has
been achieved. A sequence of simulations (not reported herein) was
performed to assess the impact of the algorithmic modifications. The
main difference introduced by the changes is an improved agreement
between numerical dissipation and heating (which can be derived from
the analysis in this work). In essence, this is an improvement in
total energy conservation. The qualitative and quantitative results
for turbulent stresses, magnetic energy, and kinetic energy are not
affected by the changes to the algorithms.

\subsection{Initial conditions}
\label{subsec:initialconditions}

The simulation starts with an analytic equilibrium disk which is
isothermal in height ($T=T(R)$, where $T$ is the temperature) and
possesses a purely toroidal net-flux magnetic field with a constant
ratio of gas-to-magnetic pressure, $\beta$. The derivation of the disk
equilibrium and a detailed description of the initial conditions can
be found in \cite{Parkin:2013}. In cylindrical coordinates ($R,z$),
the density distribution, in scaled units, is given by,
\begin{equation}
  \rho(R,z) = \rho(R,0) \exp\left(\frac{-(\Phi(R,z) - \Phi(R,0))}{T(R)} \frac{\beta}{1 + \beta} \right), \label{eqn:rho}
\end{equation}
\noindent where the pressure, $p = \rho T$, and $\beta = 2p/|B|^2
\equiv 2p/B_{\phi}^2$ is initially set to 20. The $\theta$-component
of the vector potential, $A_{\theta}$, is used to initialise the
magnetic field via ${\bf B} = \nabla \times {\bf A}$ with,
\begin{equation} 
  A_{\theta} = \frac{1}{r} \int_{r_{0}}^{r} r B_{\phi} \d r, 
\end{equation}
where $r_{0} = 7$ and $A_{\theta}(r \leq r_{0})=0$.

For the radial profiles $\rho(R,0)$ and $T(R)$ in Eq~(\ref{eqn:rho})
we use simple functions inspired by the \cite{Shakura:1973} disk
model, combined with an additional truncation of the density profile
at a specified outer radius:
\begin{eqnarray}
  \rho(R,0) & = & \rho_{0} f(R,R_{0},R_{\rm out})
  \left(\frac{R}{R_{0}}\right)^{\epsilon}, \\
  T (R) & = & T_{\rm 0} \left(\frac{R}{R_{0}}\right)^{\eta}, \label{eqn:temp}
\end{eqnarray}
\noindent where $\rho_{0}$ sets the density scale, $R_{0}$ and $R_{\rm
  out}$ are the radius of the inner and outer disk edge, respectively,
$f(R,R_{0},R_{\rm out})$ is a tapering function \citep{Parkin:2013},
and $\epsilon$ and $\eta$ set the slope of the density and temperature
profiles, respectively. Values are set to $R_{0}=7$, $R_{\rm out}=50$,
$\rho_{0}=10$, $\epsilon=-33/20$, $\eta=-9/10$, and
$T_{0}=1.5\times10^{-3}$, producing a disk with $H/R=0.1$. Note that
in contrast to the simulations in \cite{Parkin:2013, Parkin:2013b},
which focused on a finite disk mass residing within a larger
simulation domain, the adopted grid position and extent in this work
places the simulation domain {\it within} the disk
\citep[e.g.][]{Fromang:2006, Flock:2011}.

The rotational velocity of the disk is close to Keplerian, with a
minor modification due to the gas and magnetic pressure gradients,
\begin{eqnarray}
  v_{\phi}^2(R,z) = v_{\phi}^2(R,0) + (\Phi(R,z) -
  \Phi(R,0)) \frac{R}{T}\frac{d T}{d R}, \label{eqn:vphi}
\end{eqnarray}
\noindent where,
\begin{eqnarray}
v_{\phi}^2(R,0) = R \frac{\partial
  \Phi (R,0)}{\partial R} + \frac{2 T}{\beta} + \nonumber \\
 \left(\frac{1 + \beta}{\beta} \right)\left(\frac{R
    T}{\rho(R,0)}\frac{\partial \rho(R,0)}{\partial R}
 + R \frac{d T}{d R}\right).   \label{eqn:vphi0}
\end{eqnarray}
To initiate the development of turbulence in the disk, we add random
poloidal velocity fluctuations of amplitude $0.01~c_{\rm s}$, where
$c_{\rm s}$ is the sound speed, to the initial equilibrium.

\subsection{Diagnostic averages}
\label{subsec:diagnostics}

A volume-averaged value (denoted by angled brackets) for a variable
$q$ is computed via,
\begin{equation}
  \langle q\rangle = \frac{\iiint q \ts r^2 \sin \theta \d r \ts  \d
    \theta \ts \d\phi}{\iiint r^2 \sin
    \theta \ts \d r \ts \d\theta \ts d\phi}.
\end{equation}
Similarly, azimuthal averages are denoted by square brackets,
\begin{equation}
[q] = \frac{\int q r \sin\theta d\phi}{ \int r \sin \theta d\phi}.
\end{equation}
The analysis in the remainder of this paper concentrates on the ``disk
body'' which is defined as the region between $15<r<25$ and
$|z|<2H$. We also make use of time averages, all of which are computed
over the interval $20<t<40~P^{\rm orb}_{30}$, where $P^{\rm orb}_{30}$
corresponds to the orbital period at a radius, $r=30$. This time
interval corresponds to the latter half of the simulation, when a
quasi-steady state has been reached (see \S~\ref{sec:character}).

\subsection{Mean and turbulent fields}
\label{subsec:mean_turb}

In our analysis we decompose the velocity and magnetic fields into
mean and turbulent components. For this purpose we define the mean
field to be the azimuthal average of the velocity/magnetic field at a
given $r$ and $\theta$. The turbulent velocity and magnetic fields are
then given by,
\begin{equation}
\begin{array}{ c c c}
  v_{\rm i} = v'_{i} + \bar{v}_{i}, &{\rm where},
  &\bar{v}_{i} = [\rho v_{i}]/[\rho]
\end{array}
\end{equation}
and,
\begin{equation}
\begin{array}{ c c c}
  B_{i} = B'_{\rm i} + \bar{B}_{i}, &{\rm where},
  &\bar{B}_{i} = [B_{i}]
\end{array}
\end{equation}
where over-bars and primes indicate mean and turbulent components,
respectively. The corresponding mean and turbulent magnetic energies
are defined as $u_{\bar{B}}=\frac{1}{2}|\bar{B}|^2$ and
$u_{B'}=\frac{1}{2}|B'|^2$, respectively, and the turbulent kinetic
energy is given by, $u_{K'}=\frac{1}{2}\rho |v'|^2$. Note that a
mass-weighted average is used for velocities as it simplifies terms in
the mean-field equations, particularly the turbulent kinetic energy,
and also ensures that the mean flow conserves mass \citep{Favre:1969,
  Kuncic:2004}. Furthermore, for a mass-weighted average, the averaged
total kinetic energy is the sum of the mean and turbulent kinetic
energies, $[\rho v^2] = [\rho \bar{v}^2] + [\rho v'^2]$. In
\S~\ref{sec:control_volume} we perform statistical (Reynolds)
averaging for various equations, following the standard rules for
averaging \citep[see, e.g.,][]{Davidson:2004, Kuncic:2004}. For
example, $[B'_{i}] =\overline{B'_{i}}=0$, $[B'_{i}B'_{j}
\bar{v}_{i;j}]= \overline{B'_{i}B'_{j}} \bar{v}_{i;j}$, and $[B'_{i}
\bar{B}_{j}\bar{v}_{i;j}]=0$. In addition, azimuthally-averaged
equations will be converted to volume-averaged ones using an
intermediate meridional-average (i.e. over the $r,\theta$ plane):
\begin{equation}
  \frac{\iint [f]\ts r \ts\d r\ts \d \theta}{\iint r \ts\d r \ts\d \theta} \equiv
  \frac{\iiint f r^2 \ts\sin\theta \ts \d r \ts \d\theta \ts \d\phi}{\iiint  r^2
    \sin\theta \ts \d r \ts \d\theta \ts \d\phi} = \langle f
  \rangle. \label{eqn:vol_av}
\end{equation}
Note that this latter operation commutes because azimuthal averages
are computed at a specific $r$ and $\theta$. Hence, $\int r \sin
\theta \d \phi \iint \ts r \ts \d r \ts \d \theta \equiv \iiint r^2
\sin \theta \ts \d r \ts \d \theta \ts \d \phi$.
 
\begin{table}
\begin{center}
  \caption[]{Time averaged simulation diagnostics. Values are
    spatially averaged within the disk body
    (\S~\ref{subsec:diagnostics}) and time averaged over the interval
    $20 \leq t \leq 40\;P_{30}^{\rm orb}$. See \S~\ref{sec:character}
    for symbol definitions.} \label{tab:global_models}
\begin{tabular}{llll}
  \hline
Parameter & Value & Parameter & Value \\
  \hline
$N_{\rm r}$ &  0.76 & $\lb \beta_{r}\rb$ & 118 \\
$N_{\theta}$ & 0.65 & $\lb \beta_{\theta} \rb$ & 316 \\
$N_{\phi}$ &  0.81 &   $\lb \beta_{\phi} \rb$ & 20 \\
$\lb \alpha_{\rm P}  \rb$ & 0.036 & $\lb \beta_{tot} \rb$ & 16 \\
$\lb \alpha_{\rm M} \rb$ & 0.47 & $-\langle B'_{r}B'_{\phi}\rangle / \lb \rho v'_{r}v'_{\phi} \rb$& 2.69\\
  \hline
\end{tabular}
\end{center}
\end{table}

\section{Characterising the disk turbulence}
\label{sec:character}

\subsection{Turbulent stresses}

To quantify the efficiency of angular momentum transport due to the
$r-\phi$ component of the combined Reynolds and Maxwell stress, we
compute the \cite{Shakura:1973} $\alpha$-parameter
(Table~\ref{tab:global_models}),
\begin{equation}
  \langle \alpha_{P}\rangle = \frac{\langle
    \rho v'_{r}v'_{\phi} - B'_{r}B'_{\phi}\rangle}{\langle p\rangle},
\end{equation}
where we have defined the turbulent components, denoted by primes, as
described in \S~\ref{subsec:mean_turb}. This approach has been shown
above, and by \cite{Flock:2011}, to remove the influence of strong
vertical and radial gradients when computing averaged values. In
addition, we calculate the $r-\phi$ component of the Maxwell stress
normalised by the magnetic pressure,
\begin{equation}
  \langle \alpha_{\rm M}\rangle = \frac{\langle -2
    B'_{r} B'_{\phi}\rangle }{\langle |B'|^2\rangle }. \label{eqn:alphaM}
\end{equation}
The value of $\lb \alpha_{\rm P} \rb=0.036$ is in agreement with the
models within the regime of convergence reported by
\cite{Parkin:2013b}, and $\lb \alpha_{\rm M} \rb = 0.47$ agrees with
the well-resolved models presented by \cite{Parkin:2013,Parkin:2013b}
and \cite{Hawley:2013} - additional tests have confirmed that the
slightly higher value of $\lb \alpha_{\rm M} \rb$ compared to previous
work results from using the turbulent, rather than total, magnetic
fields in Eq~(\ref{eqn:alphaM}). Furthermore, the ratio of
Maxwell-to-Reynolds stress, $-\langle B'_{r}B'_{\phi}\rangle / \lb
\rho v'_{r}v'_{\phi} \rb = 2.69$ is consistent with previous
global disk studies \citep{Fromang:2006, Beckwith:2011}.

\begin{figure}
  \begin{center}
    \begin{tabular}{c}
\resizebox{80mm}{!}{\includegraphics{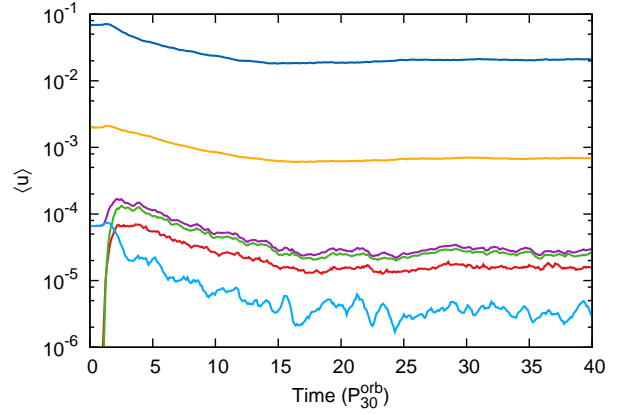}} \\
 \end{tabular}
 \caption{The time evolution of volume averaged values for various
   energies (computed over the disk body): total kinetic (dark blue),
   internal (orange), total magnetic (purple), turbulent magnetic
   (green), turbulent kinetic (red), and mean magnetic (light
   blue). Time is in units of the orbital period at a radius of
   $r=30$, $P^{\rm orb}_{30}$. (For comparison, $P^{\rm
     orb}_{30}=9~P^{\rm orb}_{8}$, therefore roughly 360 inner disk
   orbits are covered.)  The mean kinetic energy is not plotted as it
   is indistinguishable from the total kinetic energy on this
   plot. Time averaged values for energies are noted in
   Table~\ref{tab:energies}.}
    \label{fig:u_tot_turb}
  \end{center}
\end{figure}

\subsection{MRI-resolution}

To demonstrate that the simulation is sufficiently well resolved we
quote time-averaged values for the resolvability, $N_{\rm i}$, which
is defined to be the fraction of cells in the disk body that resolve
the wavelength of the fastest growing MRI mode, $\lambda_{\rm MRI-i}$,
with at least 8 cells\footnote{This equates to measuring the fraction
  of cells which have a ``quality factor'' \citep{Noble:2010,
    Hawley:2011} which is 8 or better throughout the disk - see also
  \cite{Sorathia:2012} and \cite{Parkin:2013b}.}, where,
\begin{equation}
  \lambda_{\rm MRI-i} = \frac{2 \pi |v_{\rm A i}| r \sin \theta}{v_{\phi}},
\end{equation}
where $i=r,\theta,\phi$, and $v_{\rm A i}=B_{\rm i}/\sqrt{\rho}$ is
the Alfv{\' e}n speed. The simulation achieves time-averaged values
for the resolvability of $N_{\rm r}=0.76$, $N_{\theta}=0.65$, and
$N_{\phi}=0.81$, indicating an adequate resolution of MRI modes
\citep{Sorathia:2012, Parkin:2013b}. Further discussion of the
resolution of the MRI in similar stratified global models can be found
in \cite{Fromang:2006}, \cite{Flock:2011}, \cite{Hawley:2011,
  Hawley:2013}, \cite{Parkin:2013, Parkin:2013b}, and
\cite{Suzuki:2013}.

\subsection{Mean and turbulent fields/energies}

\begin{figure*}
  \begin{center}
    \begin{tabular}{cc}
\resizebox{80mm}{!}{\includegraphics{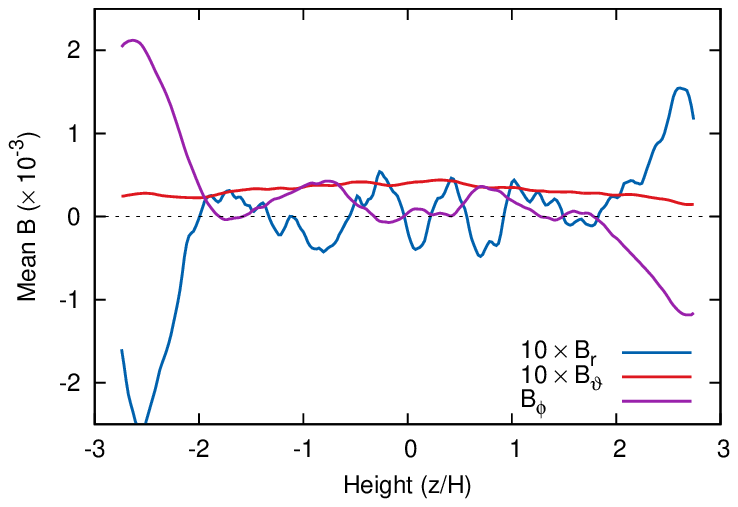}} &
\resizebox{80mm}{!}{\includegraphics{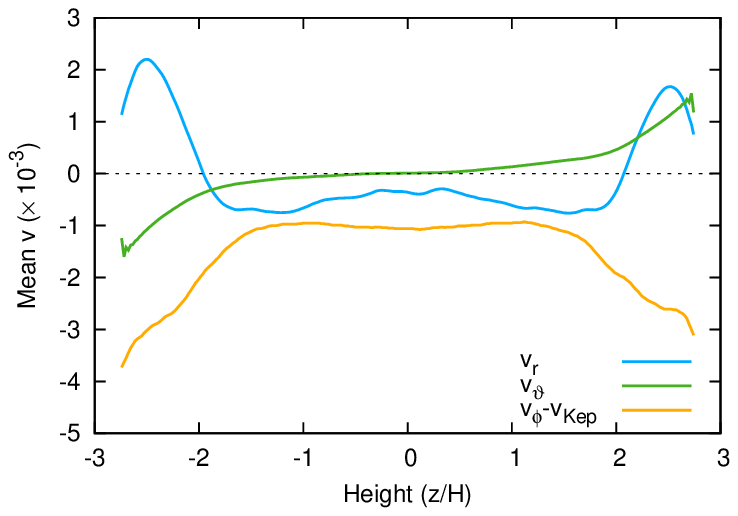}} \\
\resizebox{80mm}{!}{\includegraphics{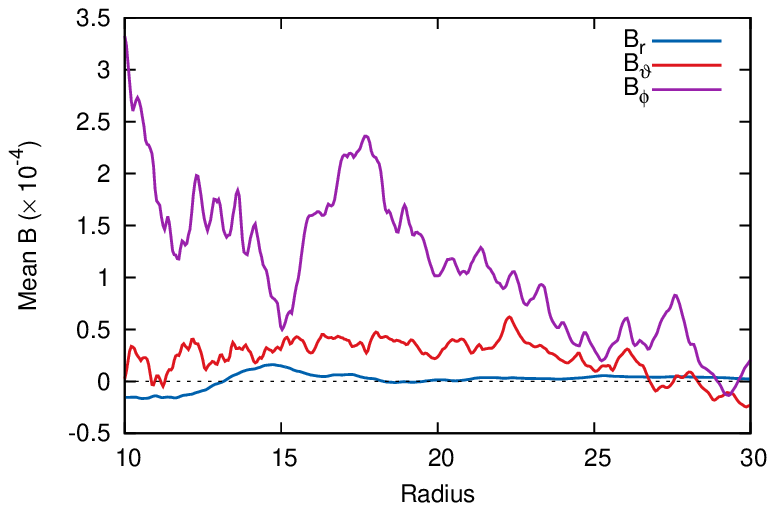}} &
\resizebox{80mm}{!}{\includegraphics{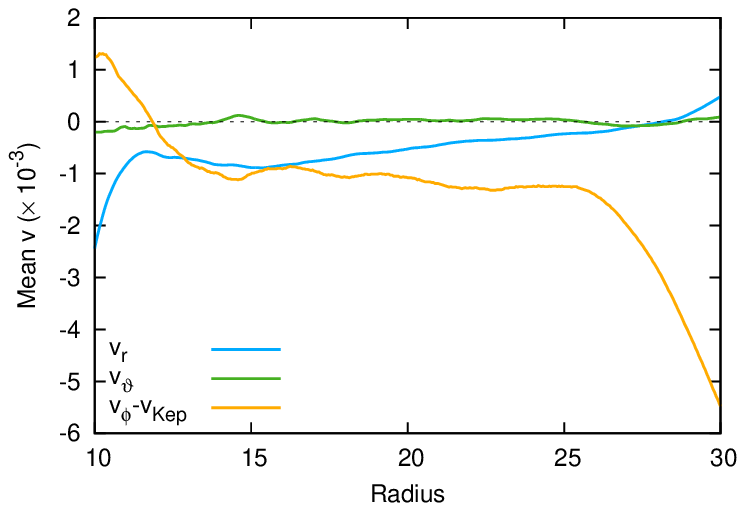}} \\
  \end{tabular}
  \caption{Time averaged profiles for mean magnetic (left column) and
    velocity (right column) fields. The top row shows vertical
    profiles, which were computed by radially averaging the azimuthal
    mean at a given $\theta$ between $15<r<25$. The lower row shows
    radial profiles, which were computed by vertically averaging
    (within $|z|<2H$) the azimuthal mean at a given radius. Time
    averages were computed over the interval $20<t<40~P^{\rm
      orb}_{30}$. In the upper left plot values of $\bar{B}_{r}$ and
    $\bar{B}_{\theta}$ have been multiplied by a factor of 10 to aid
    comparison.}
    \label{fig:av_vert}
  \end{center}
\end{figure*}

The time evolution of the various disk-body-volume-averaged energies
is shown in Fig.~\ref{fig:u_tot_turb}, with the corresponding
time-averaged values noted in Table~\ref{tab:energies}. The total
energy content of the disk is dominated by the total kinetic energy,
which is mainly due to the mean disk rotation. The internal energy is
the next largest energy, followed by the total magnetic
energy. Decomposing the kinetic and magnetic energies into mean and
turbulent components (see \S~\ref{subsec:mean_turb}), one sees that
the turbulent kinetic and magnetic energies are subthermal (i.e. less
than the internal energy). Furthermore, the turbulent kinetic energy
makes up a meagre 0.1$\%$ of the total kinetic energy, whereas more
than $90\%$ of the magnetic energy resides in the turbulent field,
with the remaining $\sim10\%$ in the mean magnetic field.

\begin{table}
\begin{center}
  \caption[]{Time averaged energies. Values are spatially averaged
    within the disk body (\S~\ref{subsec:diagnostics}) and time
    averaged over the interval $20 \leq t \leq 40\;P_{30}^{\rm
      orb}$. See \S~\ref{sec:character} for symbol definitions. See
    Fig.~\ref{fig:u_tot_turb} for the corresponding time evolution of
    these energies.} \label{tab:energies}
\begin{tabular}{llllll}
  \hline
 $\lb u_{K} \rb$ & $\lb u_{K'} \rb$ & $\lb u_{B} \rb$ & $\lb u_{B'}
 \rb$ & $\lb u_{\bar{B}} \rb$ & $\lb u_{\epsilon} \rb$ \\
 $(10^{-2})$ & $(10^{-5})$ & $(10^{-5})$ & $(10^{-5})$ & $(10^{-6})$ & $(10^{-4})$ \\
  \hline
  2.0 & 1.6 & 2.8 & 2.5 & 3.5 & 6.7 \\
\hline
\end{tabular}
\end{center}
\end{table}

\begin{figure*}
   \begin{center}
    \begin{tabular}{ccc}
\resizebox{60mm}{!}{\includegraphics{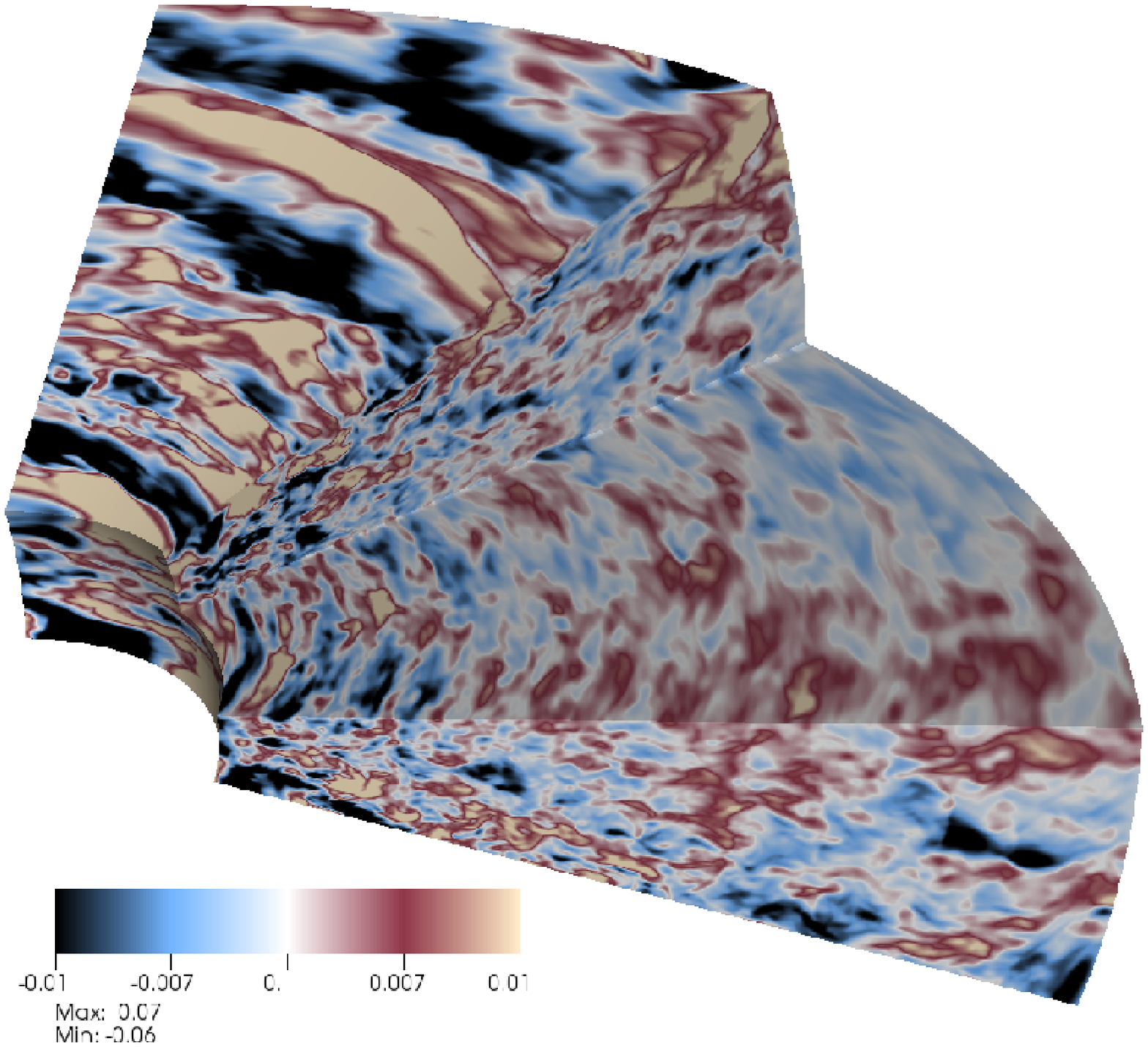}} &
\resizebox{60mm}{!}{\includegraphics{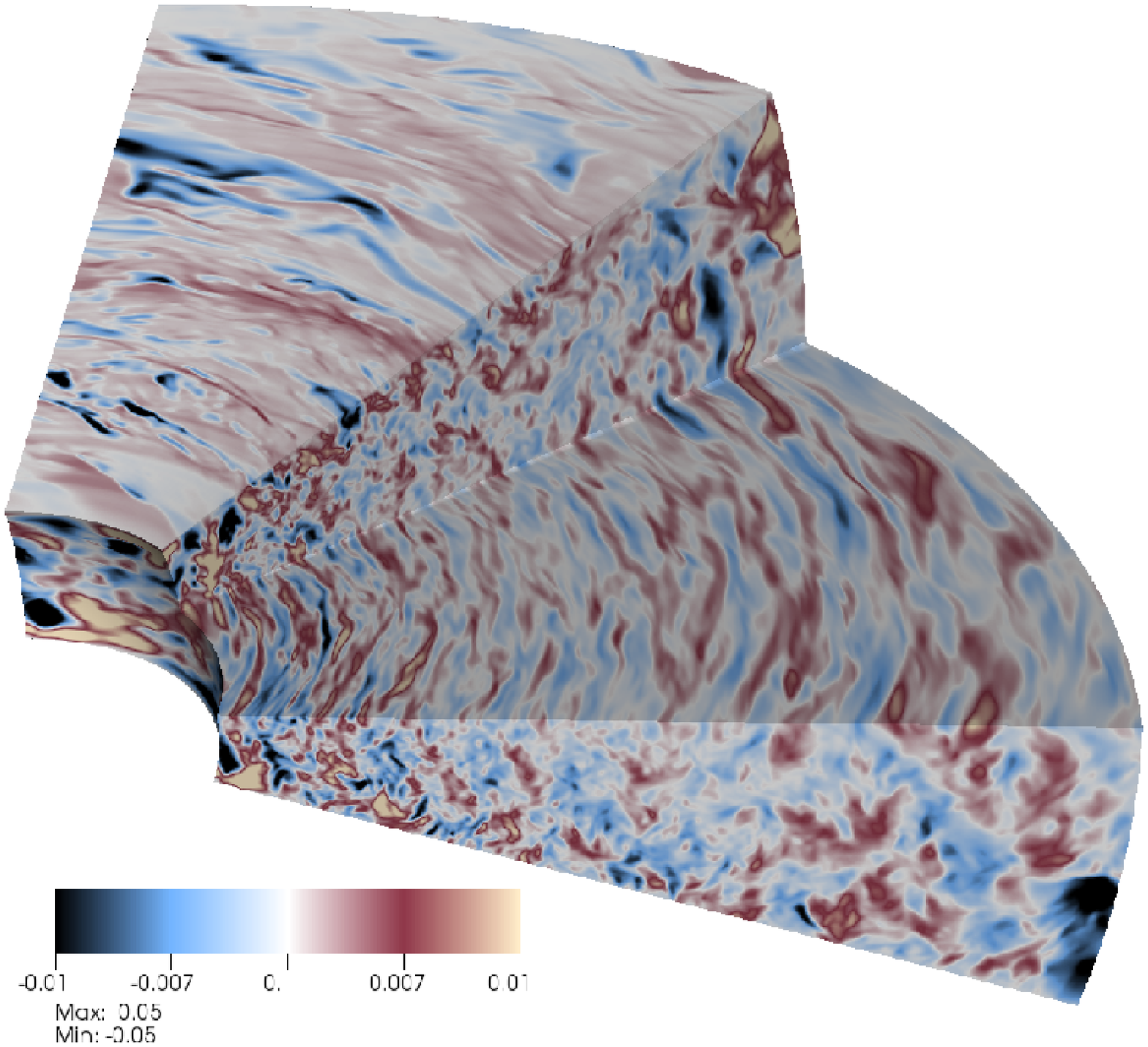}} &
\resizebox{60mm}{!}{\includegraphics{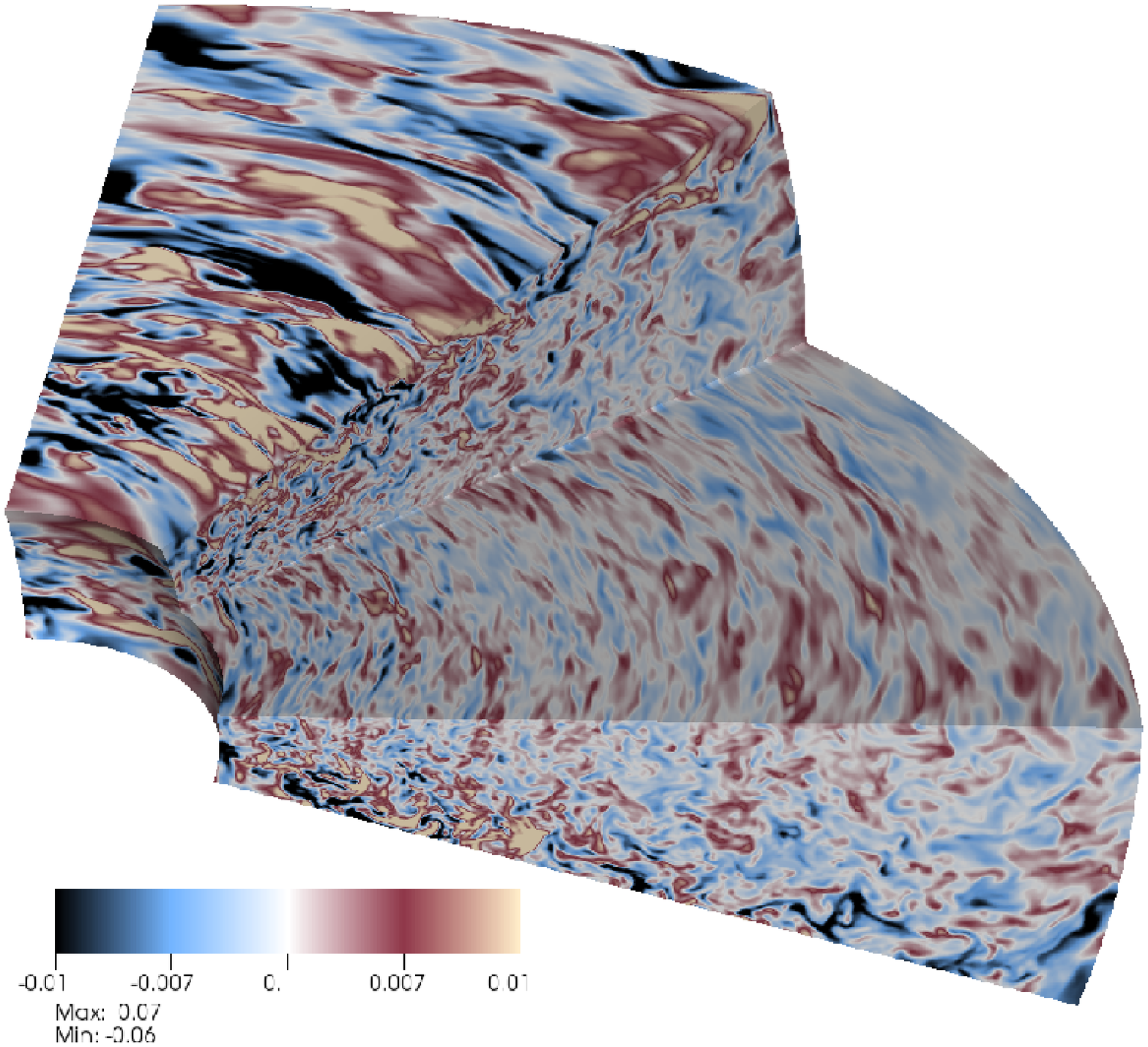}} \\
 \end{tabular}
 \caption{Snapshots of the turbulent velocity fields at $t=30~P^{\rm
     orb}_{30}$. From left to right: $v'_{r}$, $v'_{\theta}$, and
   $v'_{\phi}$. The images show a 3D view of the simulation domain
   with a section of the upper hemisphere removed, exposing the mid
   plane.}
    \label{fig:vt_snaps}
  \end{center}
\end{figure*}

Examining the vertical and radial profiles for mean magnetic fields in
more detail (Fig.~\ref{fig:av_vert}), one sees relatively little
variation of $\bar{B}_{\theta}$ as a function of height below
$|z|<2H$. There is an apparent anti-correlation between $\bar{B}_{r}$
and $\bar{B}_{\phi}$, arising from the negative shear found in
Keplerian rotation. The mean azimuthal magnetic field, on average,
increases with decreasing radius. The mean velocity field is dominated
by (almost) Keplerian rotation, where,
\begin{equation}
  v_{\rm Kep} = \left(r \frac{d \Phi}{d r} \right)^{1/2} = \frac{r^{1/2}}{r-2}, \label{eqn:vkep}
\end{equation}
with $1 - \bar{v}_{\phi} /v_{\rm Kep} \simeq 0.01$ at $r=20$. However,
Fig.~\ref{fig:av_vert} shows that at larger heights in the coronal
region ($|z|>2H$) the mean rotation is increasingly sub-Keplerian. In
the radial direction, the mean rotation transitions from sub- to
supra-Keplerian at $r\simeq13$. The mean radial flow, $\bar{v}_r$, is
of larger magnitude than the vertical flow\footnote{Note that, as
  plotted in Fig.~\ref{fig:av_vert}, a positive $\bar{v}_{\theta}$
  points away from the mid plane when $z/H$ is positive, and
  vice-versa.}, $\bar{v}_{\theta}$, and is predominantly inflowing
onto the central object. Examining the variation of $\bar{v}_r$ with
height shows that inflow velocities peak above and below the mid plane
at a height in the region of $H \leq |z| \leq 2H$. Furthermore, at
$r>28$ (which is outside of the radial extent we consider for our
``disk body'' analysis) the mean radial flow is {\it outflowing} which
is most likely the consequence of some mass carrying away angular
momentum (hence allowing accretion). There is a weak vertical outflow
at the upper/lower domain boundaries, although we are apprehensive
about designating it to be a ``wind'' as its associated mass-loss rate
is negligible compared to that arising from radial inflow within the
disk body. Furthermore, the vertical flows are substantially below the
escape velocity.

\begin{figure*}
   \begin{center}
    \begin{tabular}{ccc}
\resizebox{60mm}{!}{\includegraphics{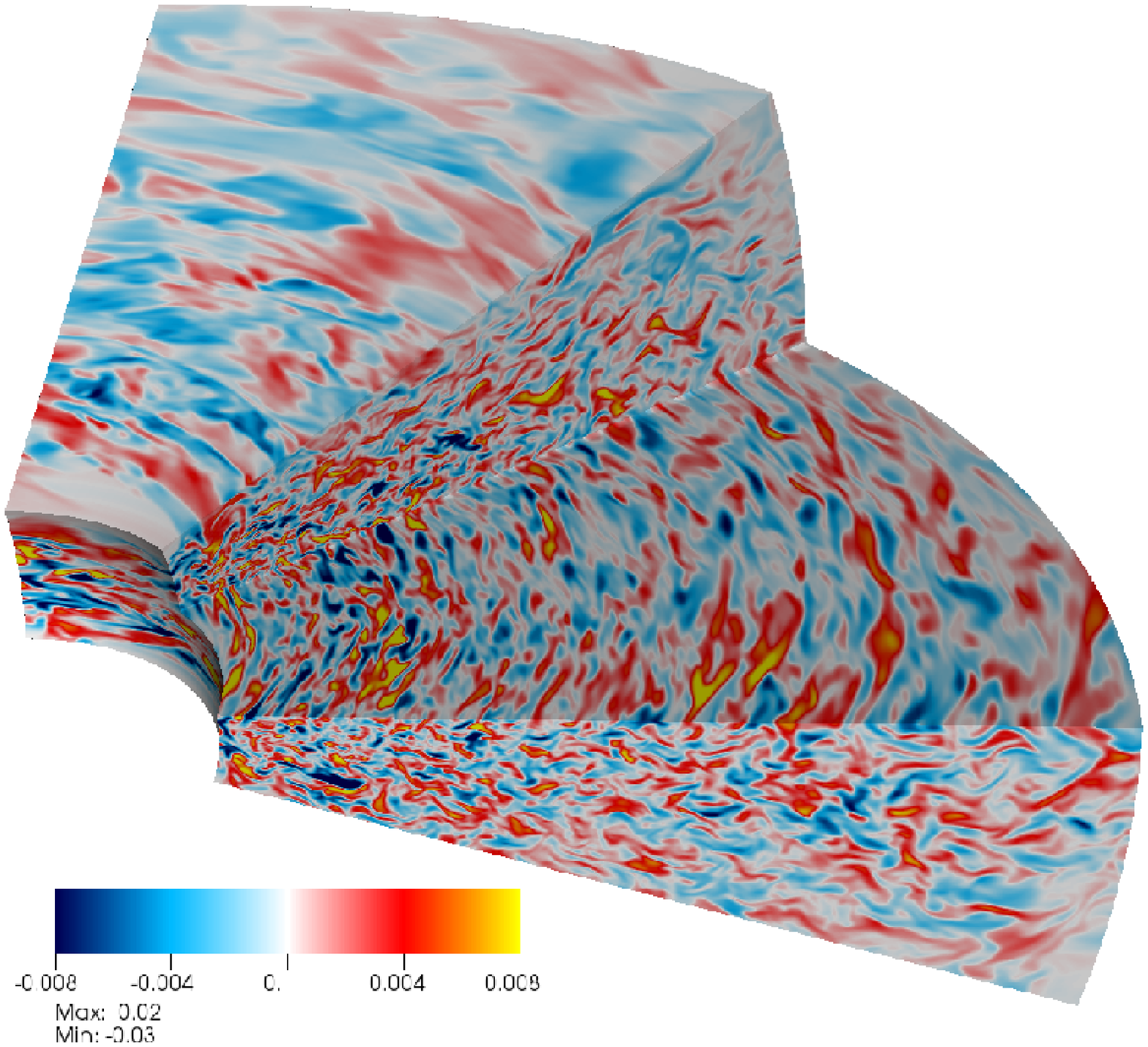}} &
\resizebox{60mm}{!}{\includegraphics{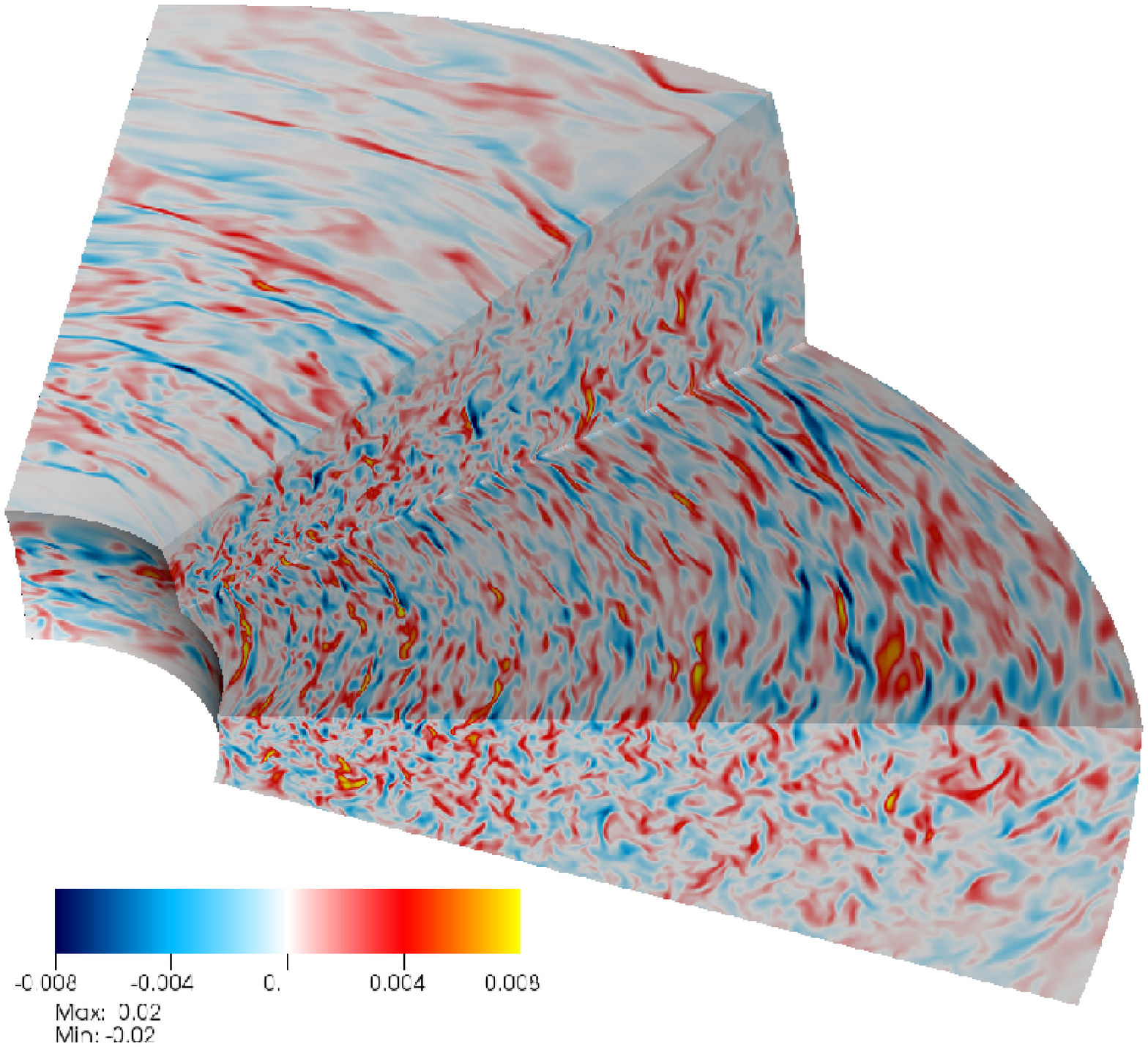}} &
\resizebox{60mm}{!}{\includegraphics{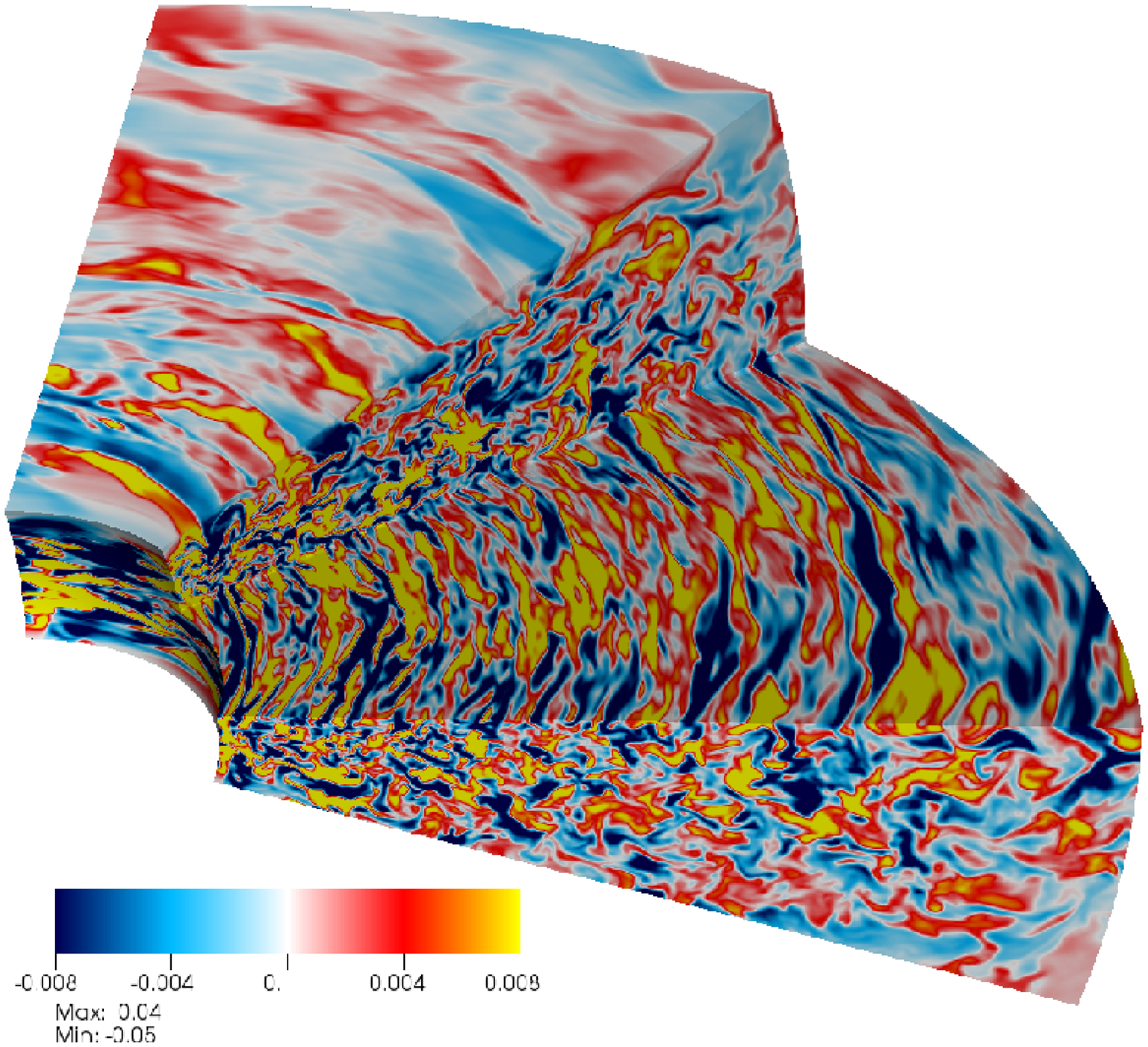}} \\
 \end{tabular}
 \caption{Same as Fig.~\ref{fig:vt_snaps} except for the turbulent
   magnetic field: $B'_{r}$ (left), $B'_{\theta}$ (middle), and
   $B'_{\phi}$ (right).}
    \label{fig:Bt_snaps}
  \end{center}
\end{figure*}

Turning next to the turbulent fields, the simulation snapshots of the
velocity and magnetic fields shown in Figs~\ref{fig:vt_snaps} and
\ref{fig:Bt_snaps}, respectively, indicate that the radial and
azimuthal fluctuations are the largest. In terms of the characteristic
shape and magnitude of the fluctuations, the turbulent velocity and
magnetic fields look notably different. However, they have common
features in the elongation of structures in the direction of rotation
and the inhomogeneous distribution of fluctuations.

\begin{figure*}
  \begin{center}
    \begin{tabular}{cc}
\resizebox{80mm}{!}{\includegraphics{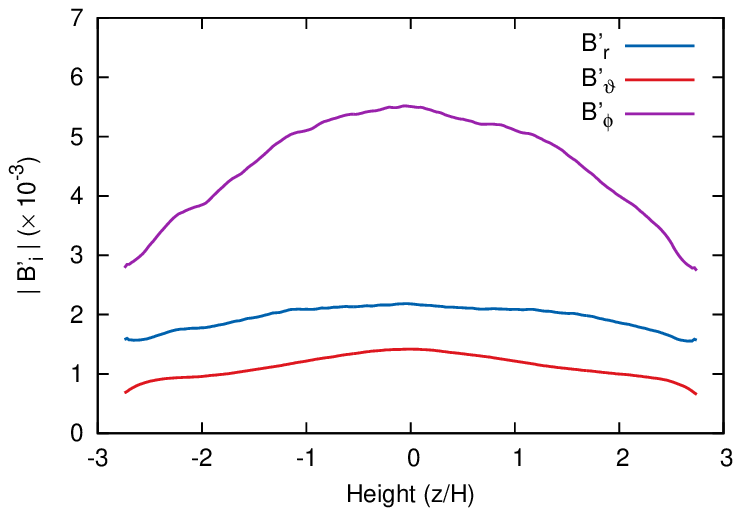}} &
\resizebox{80mm}{!}{\includegraphics{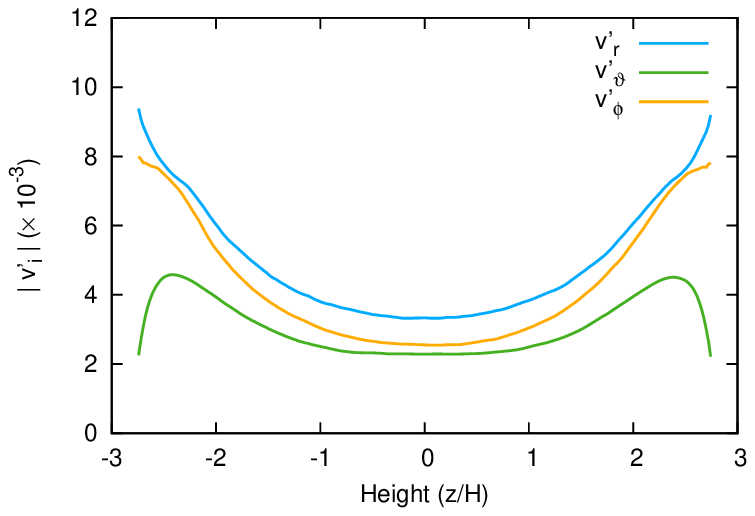}} \\
\resizebox{80mm}{!}{\includegraphics{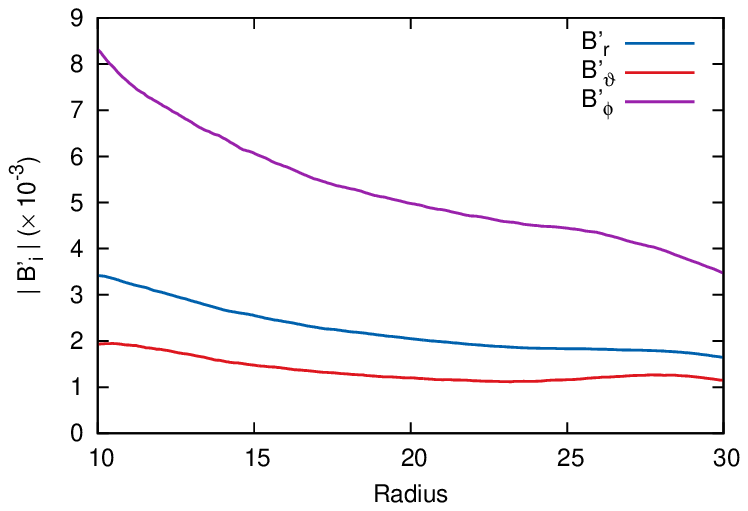}} &
\resizebox{80mm}{!}{\includegraphics{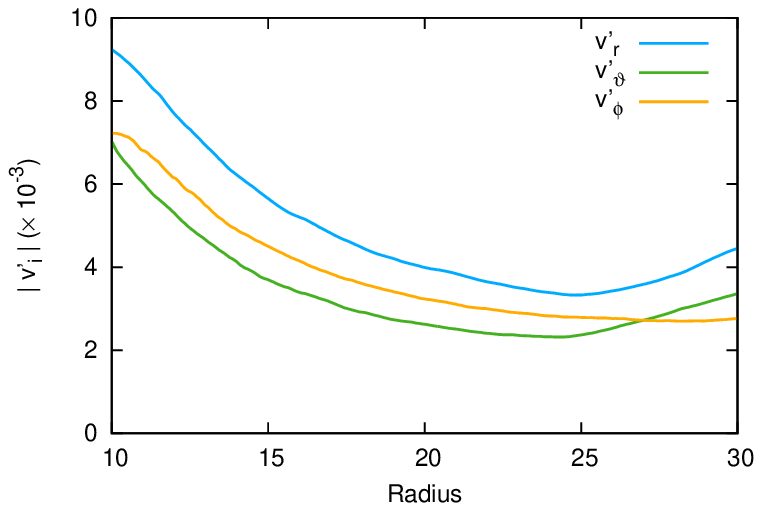}} \\
  \end{tabular}
  \caption{Time averaged profiles for the magnitude of turbulent
    magnetic (left column) and velocity (right column) fields. The top
    row shows vertical profiles, which were computed by radially and
    azimuthally averaging the turbulent field at a given $\theta$ and
    between $15<r<25$ in radius and the full domain extent
    ($0<\phi<\pi/2$) in the azimuthal direction. The lower row shows
    radial profiles, which were computed by vertically and azimuthally
    averaging the turbulent field the azimuthal mean at a given radius
    and within $|z|<2H$. Time averages were computed over the interval
    $20<t<40~P^{\rm orb}_{30}$.}
    \label{fig:abs_vert}
  \end{center}
\end{figure*}

Constructing vertical and radial profiles of the amplitude
(i.e. absolute value) of the turbulent components
(Fig.~\ref{fig:abs_vert}), shows that both magnetic and velocity
fields exhibit little variation in the region close to the mid plane
$|z|<H$, with radial velocity perturbations and azimuthal magnetic
field perturbations having the largest amplitudes. In the coronal
region ($|z|>2H$) there is anti-correlated behaviour between magnetic
and velocity field amplitudes; the amplitude of the turbulent magnetic
field falls off with height, whereas turbulent velocities grow in
amplitude with height. The turn-down in $|v'_{\theta}|$ close to the
vertical boundaries is likely due to reflected waves inhibiting
outflow\footnote{\cite{Suzuki:2009} suggest that reflected waves can
  be avoided by explicitly treating outgoing waves at boundaries using
  a characteristic decomposition.}. The rise in the amplitude of the
turbulent fields going from larger to smaller radii is the anticipated
effect of a rising Keplerian velocity at smaller radii, and thus a
larger reservoir of kinetic energy from which the MRI can drive the
turbulence. There is a rise in $|v'_r|$ and $|v'_{\theta}|$ at $r>25$
which may be associated with the outflow seen in the mean radial
velocity and the decrease in the mean disk rotation at these radii
(Fig.~\ref{fig:av_vert}). The run of turbulent velocity amplitudes
observed in Fig.~\ref{fig:abs_vert} is consistent with those found in
similar models by \cite{Fromang:2006}, \cite{Flock:2011}, and
\cite{Beckwith:2011}. Turbulent magnetic field amplitudes bear the
same ordering in height and radius as found by \cite{Flock:2011},
however we do not find a dip in the vertical profile at the mid plane
for $|B'_r|$ and $|B'_{\theta}|$, rather we observe gently peaked
profiles. 

The results simulation results provide valuable information for models
of coronal powering based on the buoyant advection of turbulent
structures \citep[e.g.][]{Blackman:2009, Uzdensky:2013}.  For
instance, the prominence of mean radial flow over mean vertical flow
rules out advection by the latter for powering coronae.  However,
close to the mid plane of the disk ($|z|<2H$), the turbulent vertical
motions are of greater magnitude than the mean radial flow. As such,
coronal powering could be realized by buoyant rising associated with
turbulent motions, provided a sufficient correlation between regions
of strong magnetic field and such motions exists.

\begin{figure}
  \begin{center}
    \begin{tabular}{c}
\resizebox{80mm}{!}{\includegraphics{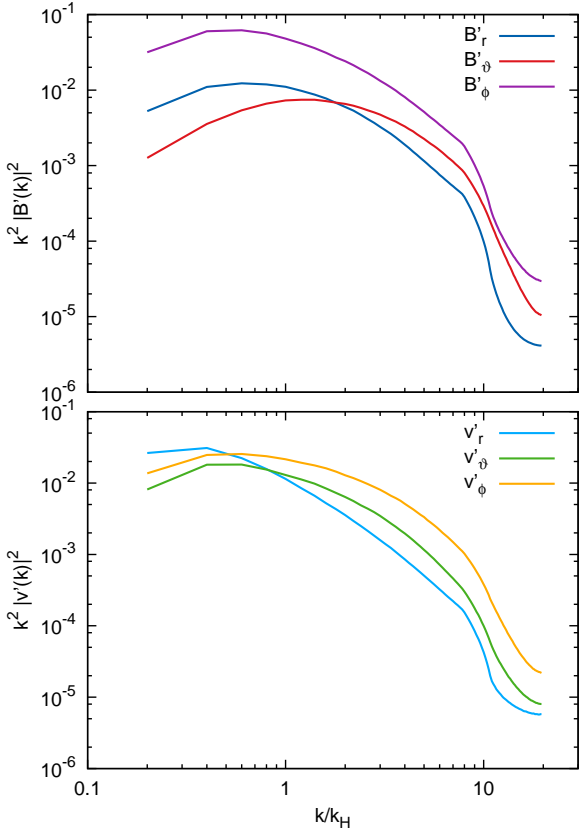}} \\
 \end{tabular}
 \caption{Angle-averaged power spectra for the turbulent magnetic
   (upper) and velocity (lower) fields, showing the separate
   directional contributions. The horizontal axis is in units of
   $k_{\rm H}=2\pi / \lb H \rb$.}
    \label{fig:Ek_dB_dv}
  \end{center}
\end{figure}

In Fig.~\ref{fig:Ek_dB_dv} we show power spectra for the radial,
vertical ($\theta$-direction), and azimuthal field components of the
turbulent magnetic and velocity fields. We define the Fourier
transform of a function $q(r,\theta,\phi)$ as,
\begin{eqnarray}
Q({\bf k}) = Q(k, \chi, \psi) = \int_0^{2 \pi} \int_0^\pi \int_0^\infty q(r,\theta,\phi) \, e^{i {\bf k \cdot x}} \times\nonumber\\
r^2 \sin \theta \, dr \, d \theta \, d \phi, \label{eqn:ft}
\end{eqnarray}
where $k$ is the radial wavenumber and $(\chi,\psi)$ are angular
coordinates in Fourier space. It then follows that the angle-averaged
(in Fourier space) amplitude spectrum,
\begin{equation}
|Q(k)|^2 = \int_0^{2 \pi} \int_0^\pi Q({\bf k}) Q^*({\bf k}) \> \sin \chi \, d\chi \, d \psi, 
\end{equation}
where an asterisk ($^*$) indicates a complex conjugate. The total
power at a given wavenumber - the power spectrum - is given by $k^2
|Q(k)|^2$. For each spectrum shown in Fig.~\ref{fig:Ek_dB_dv}, an
average was taken over 200 individual spectra computed for the disk
body region (see \S~\ref{subsec:diagnostics}) from simulation data
files equally spaced in time in the interval $20<t<40 \ts P^{\rm
  orb}_{30}$. The Fourier transforms were computed in spherical
coordinates using the method outlined in \cite{Parkin:2013b}.

The power spectra provide information about the orientation of the
fluctuating magnetic/velocity fields on different spatial scales. As
such, they allow a quantification of properties of the turbulence
suggested from a visual inspection of Figs.~\ref{fig:vt_snaps} and
\ref{fig:Bt_snaps}. Examining the upper panel of
Fig.~\ref{fig:Ek_dB_dv}, one sees that azimuthal magnetic field
dominates at all scales. However, the ordering of the radial and
vertical magnetic field components differs between the largest ($k <
k_{\rm H}$) and smallest scales ($k > k_{\rm H}$); on the largest
scales $|B'_{r}(k)|^2 > |B'_{\theta}(k)|^2$, and vice-versa on the
smallest scales. The same is true for the power spectra of the
turbulent velocity components. In fact, on the very largest scales
($k/ k_{\rm H} < 0.4$), radial velocity fluctuations are more powerful
than those for the azimuthal components, whereas close to the
dissipation scale $|v'_{r} (k)|^2$ is smaller than both
$|v'_{\theta}(k)|^2$ and $|v'_{\phi}(k)|^{2}$. The above results
indicate an important deviation from Kolmogorov-type turbulence in
that anisotropy is not removed as energy cascades to smaller
scales. To the contrary, accretion disk turbulence is anisotropic at
all scales. However, the order of strengths for different directional
contributions changes, with a possible implication that the underlying
turbulent dynamo may change character at different length scales
\citep[see also the discussion by][]{Davis:2010}.

In summary, the simulations resolve the MRI sufficiently well to
achieve a quasi-steady turbulent state in which $\langle \alpha_{\rm
  P}\rangle$ is approximately constant from $t\gtsimm 15~P^{\rm
  orb}_{30}$ onwards. The addition of a mass source term to the
simulation setup (see \S~\ref{sec:model}) allows the total disk mass
to stabilise, and this in turn produces quasi-steady energies in the
post-transient phase ($t\gtsimm 20~P^{\rm orb}_{30}$ -
Fig.~\ref{fig:u_tot_turb}). The mean velocity field is dominated by
(essentially) Keplerian rotation. There are no obvious underlying
gradients in the averaged turbulent velocity and magnetic fields,
indicating that the adopted decomposition of mean and turbulent fields
is effective.

\section{Control volume analysis}
\label{sec:control_volume}

In the previous section we provided a general qualitative view of a
turbulent accretion disk. The purpose of this section is to assess the
roles of the mean and turbulent fields in driving the energy budget of
the disk. This involves evaluating mean-field, Reynolds averaged
equations using the simulation data, and to this end we examine the
mean and turbulent magnetic energies, mean field induction equation,
turbulent kinetic energy, and the internal energy. The equations are
general. However, the analysis focuses on the disk body region
(defined in \S~\ref{subsec:diagnostics}) the boundaries of which are
open in the radial and vertical directions, and periodic in the
azimuthal direction.

\subsection{Turbulent magnetic energy evolution}
\label{subsec:mag_energy}

The derivation of the turbulent magnetic energy equation begins with
the magnetic field induction equation, to which we add a term for {\it
  numerical} resistive losses, $d^{\rm res}$, such that
Eq~(\ref{eqn:induction}) now reads,
\begin{equation}
\frac{\partial B_{\rm i}}{\partial t} = \nabla \times ({\bf v \times
  B}) + d_{\rm i}^{\rm res}. \label{eqn:induction2}
\end{equation}
The $d^{\rm res}$ term encapsulates the dissipation due to the
truncated order of accuracy of numerical finite volume codes (such as
the {\sevensize PLUTO} code used in this investigation). In essence,
$d^{\rm res}$ is a place-holder for an adopted/relevant form for the
resistive term. Next we expand the magnetic and velocity fields in
Eq~(\ref{eqn:induction2}) into mean and turbulent components (see
\S~\ref{subsec:mean_turb}), take the scalar product of $B'_{i}$ with
the resulting equation, and then azimuthally (Reynolds) average. After
a little algebra, one has,
\begin{eqnarray}
\frac{\partial \overline{u_{B'}}}{\partial t} &=& - \overline{\mathcal{P}'_{k,k}}
+ \bar{B}_{j}\overline{B'_{i} v'_{i;j}} + \overline{B'_{i}
  B'_{j}}\bar{v}_{i;j}  - 2 \frac{\partial }{\partial x_{j}} (\overline{u_{B'}}
    \bar{v}_{j})\nonumber \\ && 
- \overline{B'_{i}v'_{ j}}\bar{B}_{ i;j} - \bar{B}_{i}\overline{B'_{ i} v'_{k,k}} - \overline{v'_{i}\frac{\partial
    M'_{ij}}{\partial x_{j}} } + \overline{B'_{i} d_{i}^{\rm res}}, \label{eqn:mag_energy1}
\end{eqnarray}
where a subscript comma denotes partial differentiation, a subscript
semicolon denotes a covariant derivative, and we recall that an
over-bar indicates an azimuthal average and that $u_{B'}$ is the
turbulent magnetic energy. The Poynting vector,
\begin{equation}
\mathcal{P}_{j} = |B|^2 v_{j} - v_{i}B_{i} B_{j} = u_{B}v_{j} - v_{i} M_{ij}, \label{eqn:Poynting}
\end{equation}
and the Maxwell stress tensor,
\begin{equation}
  M_{ij} = B_{i}B_{j} -  \delta_{ij} u_{B}. \label{eqn:maxstress}
\end{equation}
Note, for example, that $\mathcal{P}'_{j}$ indicates the turbulent
Poynting vector, whereby all composite velocity and magnetic fields
are turbulent also, with the same being true for the turbulent Maxwell
stress tensor, $M'_{ij}$. The final step is to take an average of
Eq~(\ref{eqn:mag_energy1}) over a meridional ($r,\theta$) plane, thus
converting azimuthal averages into an average over a volume, $V$, with
bounding surface, $s$ - see \S~\ref{subsec:mean_turb} and
Eq~(\ref{eqn:vol_av}). The volume averaged turbulent magnetic energy
equation reads,
\begin{eqnarray}
  \lb \dot{u}_{B'}\rb &=& 
\lb B'_{i}\bar{B}_{j} v'_{i;j} \rb - \lb B'_{i} v'_{j}
  \bar{B}_{i;j} \rb - \lb B'_{i} \bar{B}_{i} v'_{k,k} \rb + \llb \bar{v}_{j} \frac{\partial
   u_{B'}}{\partial x_{j}} \rrb \nonumber\\
&& \lb F_{\mathcal{P}'} \rb + \lb \mathcal{S}_{B'}\rb + \lb
\mathcal{L}'\rb + \lb F_{u_{B'}} \rb + \lb D^{\rm res}_{B'} \rb, 
\label{eqn:mag_energy2}
\end{eqnarray}
where we have introduced the additional symbols:
\begin{eqnarray}
  \lb \dot{u}_{B'} \rb &=& \frac{\partial}{\partial t}\lb u_{B'} \rb, \label{eqn:UB}\\
  \lb F_{\mathcal{P}'} \rb &=& - \lb \mathcal{P}'_{k,k} \rb = - \frac{1}{V} \int_{s}
  \mathcal{P}'_{k} \ts ds_{k}, \\ 
  \lb F_{u_{B'}} \rb &=& - 2\llb \frac{\partial}{\partial x_{k}} (u_{B'} \bar{v}_{k}) \rrb = - \frac{2}{V} \int_{s}
  u_{B'} \bar{v}_{k} \ts ds_{k}, \\ 
  \lb \mathcal{S}_{B'} \rb &=& \lb B'_{i} B'_{j} \bar{v}_{i;j} \rb \\
\lb \mathcal{L}' \rb & = & -\llb v'_{i} \frac{\partial M'_{ij}}{\partial x_{j}} \rrb, \\
\lb D^{\rm res}_{B'} \rb &=& \lb B'_{i} d_{i}^{\rm res} \rb. \label{eqn:Dnum}
\end{eqnarray}
The symbols in equations (\ref{eqn:UB})-(\ref{eqn:Dnum}) have the
following meanings: the volume averaged rate of change of magnetic
energy is given by $\lb \dot{u}'_{B} \rb$, $\lb \mathcal{S}_{B'} \rb$
represents energy production by a stress-shear correlation between
turbulent Maxwell stresses and mean flow shear, $\lb F_{\mathcal{P}'}
\rb$ is the turbulent Poynting flux, $\lb F_{u'_{B}} \rb$ is the
advective flux of turbulent magnetic energy in/out of the disk via the
mean flow, and energy extraction due to turbulent Lorentz forces doing
work on the disk is given by $\lb \mathcal{L}' \rb$. Finally, $\lb
D^{\rm res}_{B'} \rb$ denotes resistive dissipation. Note that all
terms featuring in Eqs~(\ref{eqn:mag_energy2})-(\ref{eqn:Dnum}) can be
explicitly evaluated using the simulation data. This is one benefit of
this approach as one can evaluate the (numerical) dissipation, $D^{\rm
  res}_{B'}$, without knowledge of the functional form of the
dissipation term. Note that, in this regard, the control volume
approach bears similarities with Fourier analysis methods adopted by
\cite{Fromang:2007a} and \cite{Simon:2009}. Moreover, Parseval's
theorem equates total power in real space and Fourier space, hence the
total power in correlated terms will be the same whether one evaluates
them in real space or Fourier space.

\begin{figure}
  \begin{center}
    \begin{tabular}{c}
\resizebox{80mm}{!}{\includegraphics{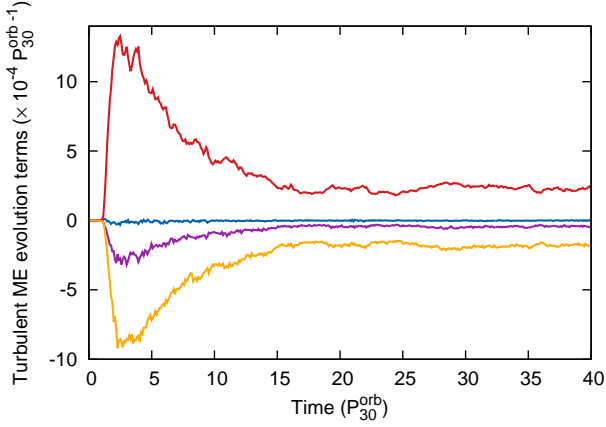}} \\
 \end{tabular}
 \caption{Turbulent magnetic energy evolution (see
   \S~\ref{subsec:mag_energy}). Shown are: $\lb F_{\mathcal{P}'} \rb$
   (blue), $\lb \mathcal{S}_{B'} \rb$ (red), $\lb \mathcal{L}' \rb$
   (purple), and $\lb D^{\rm res}_{B'} \rb$ (orange). Positive values
   indicate energy input, whereas negative values indicate removal of
   energy.}
    \label{fig:ME_control_vol}
  \end{center}
\end{figure}

Fig.~\ref{fig:ME_control_vol} shows results for the largest terms in
the turbulent magnetic energy equation (Eq~\ref{eqn:mag_energy2}) -
time-averaged values for all terms are noted in
Table~\ref{tab:control_vol}. One sees that the combination of Maxwell
stresses and shear in the mean velocity field ($\lb \mathcal{S}_{B'}
\rb$) is responsible for injecting most of the magnetic power, which
is then removed via (in this case numerical) resistive dissipation
($\lb D^{\rm res}_{B'} \rb$). This is a very similar conclusion to
that found by \cite{Parkin:2013b}, who performed a similar analysis of
the total magnetic energy equation; the present study makes the
advance of rigorously separating mean and turbulent fields.

The rate of change of turbulent magnetic energy ($\lb \dot{u}_{B'}
\rb$) has a net positive value, although the variance is larger than
the mean (Table~\ref{tab:control_vol}). The turbulent Poynting flux
($\lb F_{\mathcal{P}'} \rb$) has little impact on the energy balance,
similarly advection of turbulent magnetic energy in/out of the disk by
the mean flow ($\lb \bar{v}_{j} \frac{\partial u_{B'}}{\partial
  x_{j}}$) is relatively small (Table~\ref{tab:control_vol}). The
second largest contribution to energy removal is due to turbulent
Lorentz forces ($\lb \mathcal{L}' \rb$). We show in \S~\ref{subsec:KE}
that this term is responsible for converting turbulent magnetic energy
into turbulent kinetic energy.

It is noteworthy that the largest contribution to the turbulent
magnetic energy is not associated with the mean magnetic field, but
with the turbulent Maxwell stress and the mean velocity field
(essentially time-steady Keplerian rotation). This would suggest that,
despite exhibiting periodic oscillatory behaviour indicative of dynamo
variability \citep[\S~\ref{subsec:ind_eqn} - see also][]{Fromang:2006,
  Gressel:2010, O'Neill:2011, Oishi:2011, Flock:2012, Parkin:2013b},
mean magnetic fields do not play a significant direct part in
replenishing the turbulent magnetic energy. Hence, the role of the
mean magnetic field is not immediately evident from an analysis of the
turbulent magnetic energy alone. We return to this point in
\S\S~\ref{subsec:meanME} and \ref{subsec:ind_eqn}.

\begin{table}
\setlength{\extrarowheight}{2pt}
\begin{center}
  \caption[]{Time averaged values for terms pertaining to the control
    volume analysis. Terms are grouped (from left to right, top to
    bottom): turbulent magnetic energy, mean magnetic energy, total
    Poynting flux, turbulent Poynting flux, turbulent kinetic energy,
    and internal energy. Time averages are computed over the interval
    $20<t<40~P^{\rm orb}_{30}$.}
  \label{tab:control_vol} 
\begin{tabular}{ll  ll}
  \hline
  Term & Value & Term & Value \\
  & ($\times 10 ^{-6} P_{30}^{\rm orb \ts -1} $) &   & ($\times 10 ^{-6} P_{30}^{\rm orb \ts -1} $) \\
  \hline
\multicolumn{2}{c}{Turbulent magnetic energy} &
\multicolumn{2}{c}{Mean magnetic energy} \\
\hline 
  $\lb \dot{u}_{B'}\rb$ & $0.043\pm 0.36$ &    $\lb \dot{u}_{\bar{B}}\rb$ & $-0.066\pm0.19$ \\
  $\lb  F_{\mathcal{P}'}\rb$ & -1.1 &   $\lb \mathcal{S}_{\bar{B}} \rb$ & 14.6  \\
  $\lb  F_{u_{B'}}\rb$ & -2.5 &  $\lb \bar{B}_{i} B'_{j} v'_{i;j} \rb $ &  -1.8 \\
  $\lb  \mathcal{S}_{B'}  \rb$ & 230 &   $-\lb \bar{B}_{i} v'_{j} B'_{i;j} \rb $ &  -3.0 \\
  $\lb \mathcal{L}' \rb$ &  -43.9 &  $-\lb \bar{B}_{i} B'_{i} v'_{k,k} \rb$  & -5.2  \\
  $\lb B'_{i}\bar{B}_{j} v'_{i;j} \rb $ & -3.6 &  $ \lb F_{u_{\bar{B}}} \rb $  &  -1.1 \\
  $- \lb B'_{i} v'_{j}   \bar{B}_{i;j} \rb $ &  7.9 &  $ \llb \bar{v}_{i} \frac{\partial u_{\bar{B}}}{\partial x_{i}} \rrb$ &0.18 \\
  $-\lb B'_{i} \bar{B}_{i} v'_{k,k} \rb$ &  -5.2 &   $\lb D^{\rm
    res}_{\bar{B}}   \rb$ & -3.7 \\  
$\llb \bar{v}_{j} \frac{\partial u_{B'}}{\partial x_{j}} \rrb$  & -1.4
& & \\
 $\lb D^{\rm res}_{B'}  \rb$ & -180 & & \\
\hline 
\multicolumn{2}{c}{Poynting flux} & \multicolumn{2}{c}{Turbulent Poynting flux} \\
\hline 
$\lb  F_{\mathcal{P}}\rb$ & -27.6 &   $\lb  F_{\mathcal{P}'}\rb$ & -1.1 \\
  $\lb  F_{\mathcal{P}-r}\rb$ & -9.2 &   $\lb  F_{\mathcal{P}'-r}\rb$ & 0.1 \\
  $\lb  F_{\mathcal{P}-\theta} \rb$ & -18.4 &  $\lb  F_{\mathcal{P}'-\theta} \rb$ & -1.2 \\
  $\lb  F_{\mathcal{P}-{\rm Adv}}\rb$ & -3.3 &  $\lb  F_{\mathcal{P}'-{\rm Adv}}\rb$ & -0.5 \\
  $\lb  F_{\mathcal{P}-{\rm Stress}}\rb$ & -24.3 &  $\lb  F_{\mathcal{P}'-{\rm Stress}}\rb$ & -0.6 \\
  \hline
\multicolumn{2}{c}{Turbulent kinetic energy} & \multicolumn{2}{c}{Internal energy} \\
\hline
  $\lb \dot{u}_{K'}\rb$ & $0.041\pm0.38$ &   $\lb \dot{u}_{\epsilon} \rb $ & $3.3\pm3.1$ \\
  $\lb  \mathcal{S}_{K'}  \rb$ &  88.1 &  $ \lb F_{u_{\epsilon}} \rb $  &  -32.4 \\
  $ - \lb \mathcal{L}' \rb$  & 43.9 &  $ \lb F'_{u_{\epsilon}} \rb $ &  7.66 \\
  $ \lb F'_{K'} \rb $ & -0.5 &  $ - \lb p v'_{k,k} \rb $  & 48.1 \\
  $ \lb F_{K'} \rb$ & 0.4 &  $- \lb p \bar{v}_{k,k} \rb $ & -22.3 \\
  $\lb v'_{i} B'_{j} \bar{B}_{i;j} \rb $ & 1.9 &  $ \lb D^{\rm rad} \rb $ & -177 \\
  $ - \llb v'_{i} \frac{\partial p}{\partial x_{i}} \rrb $ & -46.4 &  $ \lb D^{\epsilon} \rb $  & 180 \\
  $ \lb D^{\rm visc} \rb$ & -87.4 & & \\ 
  \hline
\end{tabular}
\end{center}
\end{table}

\subsection{Mean magnetic energy evolution}
\label{subsec:meanME}

\begin{figure}
  \begin{center}
    \begin{tabular}{c}
\resizebox{80mm}{!}{\includegraphics{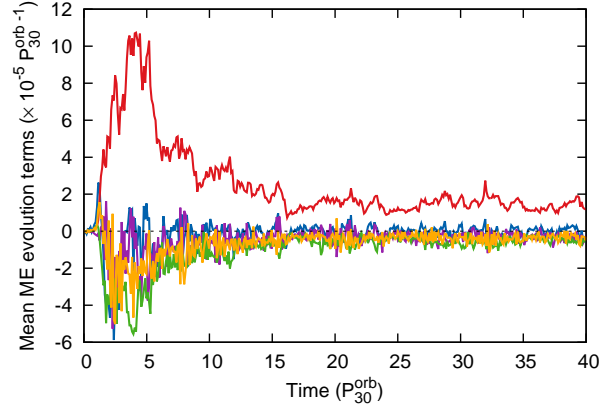}} \\
 \end{tabular}
 \caption{Mean magnetic energy evolution (see
   \S~\ref{subsec:meanME}). Shown are: $\lb \dot{u}_{\bar{B}} \rb$
   (blue), $\lb \mathcal{S}_{\bar{B}} \rb$ (red), $- \lb \bar{B}_{i}
   v'_{j} B'_{i;j} \rb$ (purple), $- \lb \bar{B}_{i} B'_{i} v'_{k,k}
   \rb$ (green), and $\lb D^{\rm res}_{\bar{B}} \rb$
   (orange). Positive values indicate energy input, whereas negative
   values indicate removal of energy.}
    \label{fig:MEm_control_vol}
  \end{center}
\end{figure}

The derivation of the mean magnetic energy equation proceeds in a
similar manner as that for the turbulent magnetic energy equation in
the previous section. The main difference is that, after expanding
fields into mean and turbulent components in
Eq~(\ref{eqn:induction2}), one takes the scalar product with
$\bar{B}_{\rm i}$ before Reynolds averaging in the azimuthal
direction, followed by a meridional ($r,\theta$) averaging. The result
is:
\begin{eqnarray}
\lb \dot{u}_{\bar{B}} \rb &=& \lb \mathcal{S}_{\bar{B}} \rb + \lb
\bar{B}_{i} B'_{j} v'_{i;j} \rb - \lb \bar{B}_{i} v'_{j} B'_{i;j} \rb
\nonumber \\ && - \lb
\bar{B}_{i} B'_{i} v'_{k,k} \rb + \lb F_{u_{\bar{B}}} \rb + \llb
\bar{v}_{i} \frac{\partial u_{\bar{B}}}{\partial x_{i}} \rrb + \lb
D^{\rm res}_{\bar{B}} \rb, \label{eqn:mean_mag_energy}
\end{eqnarray}
where we have introduced the following symbols:
\begin{eqnarray}
  \lb \dot{u}_{\bar{B}} \rb &=& \frac{\partial}{\partial t}\lb u_{\bar{B}} \rb, \label{eqn:UBm}\\
  \lb \mathcal{S}_{\bar{B}} \rb &=& \lb \bar{B}_{i} \bar{B}_{j} \bar{v}_{i;j} \rb \label{eqn:SBmean}\\
 \lb F_{u_{\bar{B}}} \rb &=& - 2\llb \frac{\partial}{\partial x_{k}} (u_{\bar{B}} \bar{v}_{k}) \rrb = - \frac{2}{V} \int_{s}
  u_{\bar{B}} \bar{v}_{k} \ts ds_{k}, \\ 
\lb D^{\rm res}_{\bar{B}} \rb &=& \lb \bar{B}_{i} d_{i}^{\rm res} \rb, \label{eqn:Dnum_m}
\end{eqnarray}
and where $u_{\bar{B}}=\frac{1}{2}|\bar{B}|^2$. The results of
evaluating Eq~(\ref{eqn:mean_mag_energy}) using the simulation data
are shown Fig.~\ref{fig:MEm_control_vol}, with time-averaged values
for all terms noted in Table~\ref{tab:control_vol}. The mean magnetic
field energy is self-generated in the sense that the dominant energy
input term is a correlation between the mean field Maxwell stress and
the mean flow shear ($\lb \mathcal{S}_{\bar{B}} \rb$ -
Eq~\ref{eqn:SBmean}). Energy is primarily removed via dissipation
($D^{\rm res}_{\bar{B}}$). It is interesting to note that terms
involving turbulent fields only appear to play a role in removing mean
magnetic energy. For instance, terms related to turbulent magnetic
field line stretching ($- \lb \bar{B}_{i} v'_{j} B'_{i;j} \rb$) and
expansions in the turbulent velocity field ($- \lb \bar{B}_{i} B'_{i}
v'_{k,k} \rb$) remove comparable amounts of mean magnetic energy to
dissipation. Compressibility is not negligible, consistent with the
findings of \cite{Gardiner:2005b}, \cite{Johansen:2009}, and
\cite{Parkin:2013b}. This feature may be linked to large scale spiral
density waves \citep[e.g.][]{Heinemann_Papaloizou:2009,
  Heinemann_Papaloizou:2012} which can be seen in snapshots of the
simulation.

In summary, based on an analysis of the turbulent
(\S~\ref{subsec:mag_energy}) and mean magnetic energies, the main
input to turbulent magnetic energy comes from correlations between
turbulent terms. Moreover, terms featuring mean-turbulent field
correlations remove mean magnetic energy, thus acting like a turbulent
resistivity. The interaction between mean and turbulent fields, and
the resulting influence on the evolution and maintenance of turbulence
in the disk, are only weakly apparent. In the following section we
turn to the mean field induction equation for further details of
mean-turbulent field interactions.

\subsection{Mean field induction equation}
\label{subsec:ind_eqn}

In our analysis, the turbulent magnetic fields are deviations from
time-dependent mean magnetic fields. Hence, although mean magnetic
fields do not feature in the largest terms in the turbulent magnetic
energy equation, their importance is implicit in the time-dependent
evolution of the turbulent magnetic field.  In this section we examine
the mean magnetic field induction equation, which exhibits a clear
interplay between mean and turbulent fields.

\subsubsection{Volume averaged form}
\label{subsubsec:ind_vol}

\begin{figure}
  \begin{center}
    \begin{tabular}{c}
\resizebox{80mm}{!}{\includegraphics{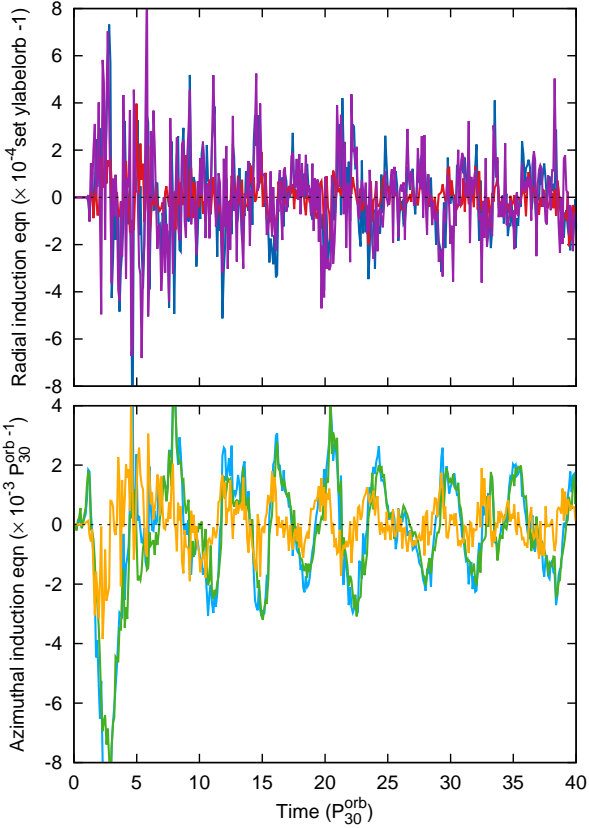}} \\
 \end{tabular}
 \caption{Volume integrated mean field induction equations for $\lb
   B_{r} \rb$ (upper) and $\lb B_{\phi} \rb$ (lower). Shown are:
   $\frac{\partial}{\partial t} \lb B_{r} \rb$ (dark blue), $-\lb
   (\nabla \times \bar{\mathcal{E}})_{r} \rb$ (red), $-\lb (\nabla
   \times \mathcal{E}')_{r} \rb$ (purple), $\frac{\partial}{\partial
     t} \lb B_{\phi} \rb$ (light blue), $-\lb (\nabla \times
   \bar{\mathcal{E}})_{\phi} \rb$ (green), and $-\lb (\nabla \times
   \mathcal{E}')_{\phi} \rb$ (orange).}
    \label{fig:ind_vol}
  \end{center}
\end{figure}

The volume-averaged mean field induction equation is, 
\begin{equation}
  \frac{\partial}{\partial t} \lb B_{i} \rb = -\lb (\nabla \times
  \bar{\mathcal{E}})_{i} \rb -\lb (\nabla \times \mathcal{E}')_{i} \rb \label{eqn:ind_vol}
\end{equation}
where $\bar{\mathcal{E}}_{i}= \epsilon_{ijk} \bar{v}_{j} \bar{B}_{k}$
and $\mathcal{E}'_{i} = \epsilon_{ijk} v'_{j} B'_{k}$ are the mean and
turbulent electromotive force (EMF), respectively. Note that we have
neglected to include a numerical resistive term in this analysis -
there is little indication from the results that it plays a
considerable role in magnetic field induction.

From Fig.~\ref{fig:ind_vol} one can see that the contributions to mean
field evolution are complex. The results are, however, suggestive of
quasi-periodic oscillations, and more so for the azimuthal induction
equation compared to the radial equation. From the upper panel of
Fig.~\ref{fig:ind_vol} the evolution of $\lb B_{r} \rb$ is more
strongly correlated with the turbulent EMF, with a secondary influence
from the mean EMF. The weak influence of the mean EMF on
$\frac{\partial}{\partial t} \lb B_{r} \rb$ arises from our azimuthal
averaging of the mean fields, which defines them as axisymmetric,
making,
\begin{equation}
  \lb (\nabla \times \bar{\mathcal{E}})_{r} \rb = \frac{1}{r \sin \theta}\frac{\partial
    \bar{\mathcal{E}}_{\phi}}{\partial \theta}, 
\end{equation}
where $\bar{\mathcal{E}}_{\phi} = \bar{v}_{\theta} \bar{B}_r -
\bar{v}_r \bar{B}_{\theta}$.  Hence, because $\bar{B}_{\theta}$ and
$\bar{v}_{\theta}$ are relatively small, and vertical gradients in
these quantities are reasonably flat below $|z|<2H$
(Fig.~\ref{fig:av_vert}), one finds $\lb (\nabla \times
\bar{\mathcal{E}})_{r} \rb$ to be small. In contrast, for $\lb
B_{\phi} \rb$ (lower panel) the mean EMF is the main driver with a
minor contribution from the turbulent EMF. In both cases, the weaker
contributor (i.e. the mean EMF in the upper panel) is slightly out of
phase, leading by roughly a quarter of an oscillation period.

\subsubsection{Induction at surfaces}
\label{subsubsec:ind_surf}

The induction equation can also provide information about the role of
different surfaces in the disk for mean magnetic field generation
\citep{Parkin:2013b}. We write the azimuthal equation in contra-variant
form as this exposes the importance of different surfaces for mean
field generation. For the mean radial and azimuthal magnetic fields,
with $\mathcal{E} = \bar{\mathcal{E}} + \mathcal{E}'$, one has,
\begin{eqnarray}
\frac{\partial}{\partial t} \lb B_{r} \rb = 
 -\frac{1}{V}\int_{\theta_2} \mathcal{E}_{\phi} \ts \d s_{\theta} 
+ \frac{1}{V} \int_{\theta_1} \mathcal{E}_{\phi} \ts \d s_{\theta},   \label{eqn:br_surf}
\end{eqnarray}
and,
\begin{eqnarray}
\frac{\partial}{\partial t} \llb \frac{B_{\phi}}{r \sin \theta} \rrb  = 
 -\frac{1}{V}\int_{r_2} \frac{\mathcal{E}_{\theta}}{r \sin \theta} \d s_{r} 
+ \frac{1}{V} \int_{r_1} \frac{\mathcal{E}_{\theta}}{r \sin \theta} \d s_{r}  \nonumber \\
-\frac{1}{V}\int_{\theta_2} \frac{\mathcal{E}_{r}}{r \sin \theta} \d s_{\theta} 
+ \frac{1}{V} \int_{\theta_1} \frac{\mathcal{E}_{r}}{r \sin \theta}\d s_{\theta}, \label{eqn:bphi_surf}
\end{eqnarray}
where $V = \iiint dV$ is the volume bound by the surfaces, $\d s_{r} =
r^2 \sin \theta \ts \d \theta \ts \d \phi$, and $\d s_{\theta} = r
\sin \theta \ts \d r \ts \d \phi$. 

Only the vertical boundaries contribute to the evolution of $\lb B_{r}
\rb$; the contribution from $\partial \mathcal{E}'_{\theta}/ \partial
\phi$ drops out due to periodicity in the azimuthal direction. Hence,
the disk-corona interface plays an intimate part in the generation of
mean radial fields in the disk. This shows that the vertical component
of the EMF cannot aid mean radial field evolution on the largest
realizable scales\footnote{\cite{Davis:2010} do, however, find
  evidence for $\partial \mathcal{E}'_{\theta} / \partial \phi$ being
  important for a small scale dynamo with the implication that
  non-axisymmetric perturbations drive energy evolution on the
  smallest scales.} \citep[see also the discussion
in][]{Brandenburg:1995, Davis:2010, Gressel:2010}. Recalling that the
turbulent EMF provides the dominant contribution to the evolution of
$\lb B_{r} \rb$ (Fig.~\ref{fig:ind_vol}), there is a suggestion that
buoyantly rising turbulent motions are crucial to mean radial field
evolution \citep[e.g.][]{Tout:1992}. Previous studies have alluded to
a crucial importance for the corona in driving the turbulence within
the disk \citep[e.g.][]{Sorathia:2010, Guan:2011, Beckwith:2011}. The
analysis in this subsection provides a natural explanation.

\begin{figure}
  \begin{center}
    \begin{tabular}{c}
\resizebox{80mm}{!}{\includegraphics{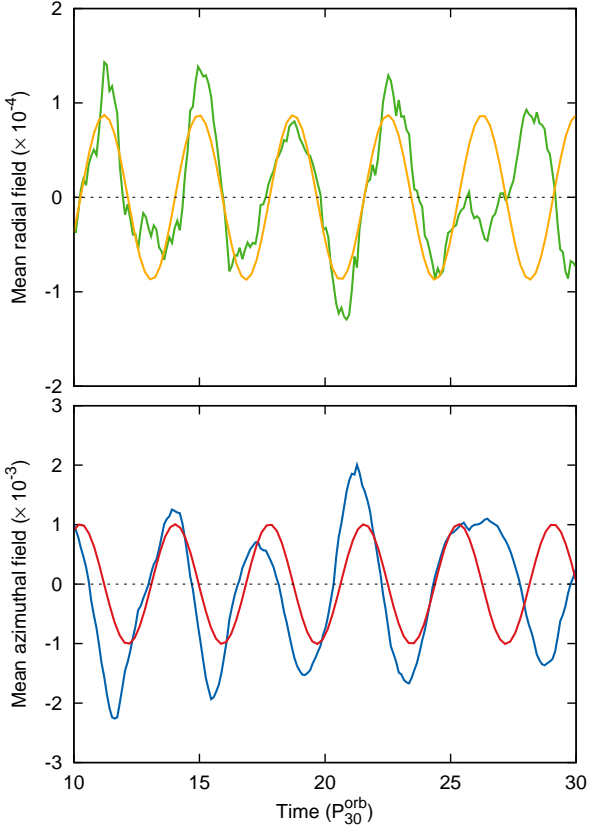}} \\
 \end{tabular}
 \caption{Comparing the basic dynamo model to the simulation data for
   the radial (upper panel) and azimuthal (lower panel) magnetic
   fields (see \S~\ref{subsubsec:dynamo_model}). Shown are: $\lb B_{r}
   \rb$ from the simulation data (green), the model $\lb B_{r} \rb$
   given by Eq~(\ref{eqn:model_Br}) (orange), $\lb B_{\phi} \rb$ from
   the simulation data (blue), and the model $\lb B_{\phi} \rb$ given
   by Eq~(\ref{eqn:model_Bphi}) (red).}
    \label{fig:dynamo_model}
  \end{center}
\end{figure}

\subsubsection{A simple harmonic oscillator dynamo model}
\label{subsubsec:dynamo_model}

Magnetic field oscillations indicative of a mean field dynamo have
been observed in a number of simulation studies of stratified disks
\citep[e.g.][]{Brandenburg:1995, Stone:1996, Miller:2000, Arlt:2001,
  Fromang:2006, Davis:2010, Gressel:2010, Shi:2010, Guan:2011,
  Simon:2011, Oishi:2011, O'Neill:2011, Flock:2012,
  Parkin:2013b}. Inspecting the data from the simulation presented in
this work, there is a striking correlation between
$\frac{\partial}{\partial t} \lb B_{r} \rb$ with $\lb B_{\phi} \rb$,
and $\frac{\partial}{\partial t} \lb B_{\phi} \rb$ with $\lb B_{r}
\rb$. \cite{Brandenburg:1995} observed similar behaviour, which led
them to search for correlations between $\lb \mathcal{E}'_{\phi} \rb$
and $\lb B_{\phi} \rb$ as, since in the context of mean field dynamo
theory \citep{Krause:1980}, that would close a simple model for mean
field evolution. \cite{Flock:2012} have repeated this exercise for
global simulations, finding the same correlation. Indeed, our
simulation results also exhibit an apparent $\lb \mathcal{E}'_{\phi}
\rb \propto \lb B_{\phi} \rb$ relation. The physical picture indicated
by these relations is the one of creation of mean toroidal field from
radial field via shear, combined with vertical gradients in $\lb
\mathcal{E}'_{\phi} \rb $ converting toroidal field into radial
field. The latter mechanism is indirect in the sense that $\lb
B_{\phi} \rb$ does not explicitly feature in $\lb \mathcal{E}'_{\phi}
\rb$. Hence, an intermediate agent, related to the turbulent fields
and in the vertical direction, must operate. Buoyant motions
\citep[e.g. Parker instability -][]{Tout:1992} and/or the
non-axisymmetric MRI \citep{BH92, Terquem:1996} seem viable. These
points are discussed further by, for example, \cite{Brandenburg:1995},
\cite{Brandenburg:2005}, \cite{Davis:2010}, \cite{Gressel:2010},
\cite{Oishi:2011}, and \cite{Flock:2012}.

In the following we consider the basic dynamo model outlined above,
which we cast as:
\begin{equation}
  \frac{\partial}{\partial t} \lb B_r \rb = \frac{1}{\tau_{\rm turb}}
  \lb B_{\phi} \rb, \label{eqn:Br_dynamo}
\end{equation}
and,
\begin{equation}
  \frac{\partial}{\partial t} \lb B_{\phi} \rb = \llb
  \frac{\d \Omega}{\d \ln r} \rrb \lb B_{r} \rb, \label{eqn:Bphi_dynamo}
\end{equation}
where $\llb \d \Omega / \d \ln r \rrb$ is the mean shear in the disk
and $\tau_{\rm turb}$ is a timescale related to an undisclosed
physical mechanism which allows the turbulence to regenerate radial
field from toroidal field. Equations~(\ref{eqn:Br_dynamo}) and
(\ref{eqn:Bphi_dynamo}) describe the mean field evolution as a simple
harmonic oscillator, affording the solution:
\begin{equation}
  \lb B_r \rb = -\omega  \llb \frac{\d \Omega}{\d \ln r} \rrb^{-1} \lb
  B_{\phi } \rb_{0} \sin \omega t + \lb B_{r} \rb_{0} \cos \omega t, \label{eqn:model_Br}
\end{equation}
and,
\begin{equation}
  \lb B_{\phi} \rb = \lb  B_{\phi} \rb_{0} \cos \omega t + \llb
  \frac{\d \Omega}{\d \ln r} \rrb \omega^{-1} \lb B_{r} \rb_{0} \sin
  \omega t, \label{eqn:model_Bphi}
\end{equation}
where $\omega = \left({- \ts \tau_{\rm turb}^{-1} \llb \d \Omega / \d
    \ln r \rrb} \right)^{1/2}$ is the oscillation frequency, and $\lb
B_{r} \rb_{0} $ and $\lb B_{\phi} \rb_{0} $ are the volume averaged
radial and azimuthal magnetic fields at time $t=0$. Note that $\omega$
is real so long as the angular velocity decreases as a function of
radius. This is satisfied in a Keplerian disk and is a pre-requisite
for the MRI \citep{BH91}. Furthermore, the dynamo period,
\begin{equation}
  \tau_{\rm dyn} = \frac{2 \pi}{\omega} = 2 \pi \left( \frac{-1}{\tau_{\rm
        turb}} \llb \frac{\d
        \Omega}{\d \ln r} \rrb   \right)^{-1/2}, \label{eqn:tdyn}
\end{equation}
is a function of the mean shear in the disk, $\llb \d \Omega /\d \ln r
\rrb$, and the turbulent timescale, $\tau_{\rm turb}$. Comparing the
basic model mean fields against the simulation data shows good
agreement (Fig.~\ref{fig:dynamo_model}), with $\tau_{\rm dyn} \simeq
3.8 \ts P^{\rm orb}_{30}$. The corresponding turbulent timescale,
$\tau_{\rm turb} \simeq 7 \ts P^{\rm orb}_{30}$, is consistent with
the growth time for a low wavenumber non-axisymmetric MRI mode seeded
from a (mean) toroidal magnetic field \citep{BH92, Parkin:2013}. The
model mean fields begin to diverge from the simulation data after an
interval of roughly $15 \ts P^{\rm orb}_{30}$. This likely arises as a
result of additional physics not included in Eqs~(\ref{eqn:Br_dynamo})
and (\ref{eqn:Bphi_dynamo}). For example, a weaker intermittent
secondary dynamo connected with the longer timescale oscillations in
$\lb B_{\theta} \rb$ \citep[see figure 11 of][]{Parkin:2013b}.

In summary, the induction equation provides insight into turbulent
fields influencing mean field evolution. A prime site for grasping
this interaction is in the mechanism(s) which allow the turbulent
fields to regenerate mean radial field from, presumably, mean toroidal
field. In the analysis above we have encapsulated this, potentially
vast, complexity in a single parameter, $\tau_{\rm turb}$. Ultimately,
unlocking the underlying physics requires a more complete closure
model \citep[e.g.][]{Ogilvie:2003, Pessah:2006, Lesur_Ogilvie:2008b}
and/or a derivation of tensorial dynamo coefficients
\citep{Brandenburg:2005, Gressel:2010}.

\begin{figure}
   \begin{center}
    \begin{tabular}{c}
\resizebox{80mm}{!}{\includegraphics{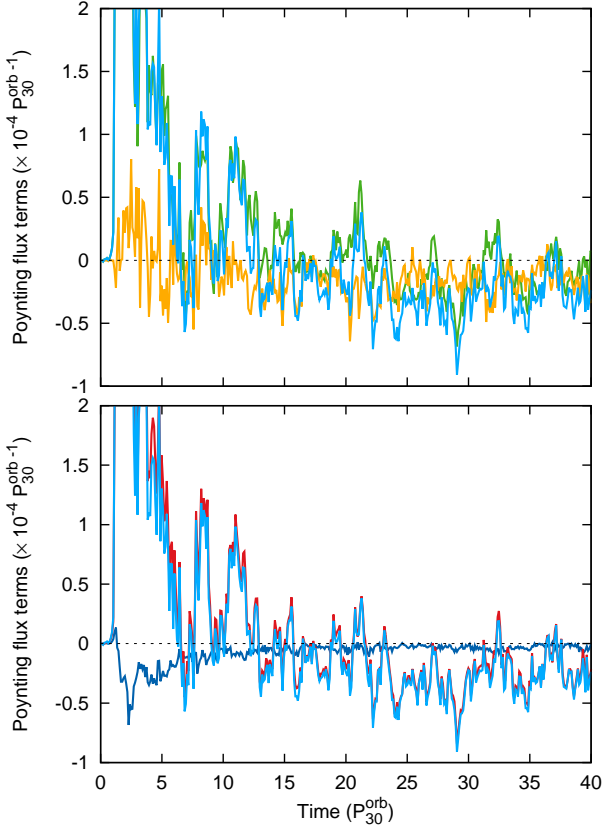}} \\
  \end{tabular}
 \caption{The Poynting flux evaluated from the simulation data (see
   \S~\ref{subsec:Poynting}). The analysis was performed over the
   region $15<r<25$ and $|z|\leq 2H$. In the upper panel the radial
   ($\lb F_{\mathcal{P} - r} \rb$ - green), vertical ($\lb
   F_{\mathcal{P} - \theta}\rb $ - orange), and total ($\lb
   F_{\mathcal{P}} \rb$ - light blue) contributions to the Poynting
   flux are plotted. The lower panel shows the contributions from the
   stress-related ($\lb F_{\mathcal{P}-{\rm Stress}} \rb$ - red) and
   advection-related ($\lb F_{\mathcal{P}-{\rm Adv}} \rb$ - dark blue)
   parts of the Poynting flux. Positive values indicate an influx of
   magnetic energy into the disk, and vice-versa for negative
   values. Corresponding time-averaged values are noted in
   Table~\ref{tab:control_vol}.}
    \label{fig:Poynting}
  \end{center}
\end{figure}

\subsection{Poynting flux}
\label{subsec:Poynting}

Models of Seyfert galaxy coronae typically invoke the presence of a
hot coronal region overlying a cool disk \citep[commonly known as
``reflection'' models -][]{Haardt:1991, Haardt:1993, Field:1993}. An
important ingredient in these models is the fraction of energy
transported to, and dissipated within, the coronal region. Hence,
knowledge of the means by which a turbulent accretion disk transports
energy into the coronal region is pivotal for accurate modelling.

The transport of electromagnetic energy through the boundaries of the
accretion disk can be assessed using the Poynting flux. Our main focus
will be on the directional contributions, and later, the
stress-related and advection-related components of the total Poynting
flux. Taking the scalar product of $B_{i}$ with the induction equation
(written in a conservative form), then averaging the resulting
equation over a volume, $V$, with bounding surface, $s$, one has:
\begin{equation}
  \frac{\partial}{\partial t} \lb u_{B} \rb = \lb F_{\mathcal{P}} \rb + \lb
  (B_{i}v_{j}- v_{i} B_{j})B_{i;j} \rb, 
\end{equation}
where the Poynting vector, $\mathcal{P}$, is given by
Eq~(\ref{eqn:Poynting}), and the volume-averaged total Poynting flux
is,
\begin{equation}
 \lb F_{\mathcal{P}} \rb = -\frac{1}{V} \int \nabla \cdot \mathcal{P} \ts \d V =
 -\frac{1}{V} \int_{s} \mathcal{P}_{j} \ts \d s_{j}.
\end{equation}
Fig.~\ref{fig:Poynting} shows the evolution of the Poynting flux in
the simulation. During the initial transient evolution of the
simulation, there is a large influx of magnetic energy in the radial
direction. As the evolution of the disk proceeds, the positive
contribution from the radial boundaries subsides and the total
Poynting flux becomes dominated by an outwardly directed flux through
the vertical boundaries (Table~\ref{tab:control_vol}). Thus, once the
quasi-steady turbulent state is reached, the Poynting flux mostly acts
to transfer magnetic energy from the disk body ($|z| \leq 2H$) into
the coronal region ($|z|>2H$).

The contribution to the Poynting flux from advection of
electromagnetic energy is small. One can see this by considering the
second equality for the Poynting vector in Eq~(\ref{eqn:Poynting}),
where the separate contributions due to advection of magnetic energy
($\propto u_{B} v_{j}$) and Maxwell stresses ($\propto v_{i} M_{ij}$)
are more clear. Defining,
\begin{equation}
\lb F_{\mathcal{P}-{\rm Adv}} \rb = -\frac{1}{V} \int_{s} u_{B} v_{j} \ts \d s_{j},
\end{equation}
and,
\begin{equation}
\lb F_{\mathcal{P}-{\rm Stress}} \rb = -\frac{1}{V} \int_{s} v_{i} M_{ij} \ts \d s_{j},
\end{equation}
it is evident from the lower panel of Fig.~\ref{fig:Poynting} that the
majority of magnetic energy is carried away by the stress-related part
of the total Poynting flux. 

Shearing-box studies of wind-launching by MRI-active disks
indicate\footnote{However, many properties of the wind-launching
  follow the standard picture of magnetocentrifugal winds
  \citep{Blandford:1982, Ogilvie:2012, Fromang:2013, Lesur:2013,
    Bai:2013}.}  that it is the conversion of turbulent magnetic
energy (carried by the turbulent Poynting flux) into kinetic energy
which initiates the outflow \citep{Suzuki:2009, Suzuki:2010,
  Io:2013}. In this case the advection-related and stress-related
components of the {\it turbulent} Poynting flux are found to be
comparable. We have not witnessed wind launching in the simulation
presented in this work. This does not seem to be due to the simulation
transporting an insufficient amount of energy into the coronae via the
turbulent Poynting flux as values for $\lb F_{\mathcal{P}'} \rb$ (and
its various components) noted in Table~\ref{tab:control_vol} are in
good agreement with recent global simulations by \cite{Suzuki:2013},
namely $\lb F_{\mathcal{P}'-{\rm Adv}} \rb \sim \lb
F_{\mathcal{P}'-{\rm Stress}} \rb $. (It should, however, be noted
that the turbulent Poynting flux is significantly smaller than the
total Poynting flux, $|\lb F_{\mathcal{P}'} \rb| \ll |\lb
F_{\mathcal{P}} \rb |$.) Possible explanations for the lack of wind
launching in the simulation are an insufficient vertical extent to the
domain or, as discussed in \S~\ref{sec:character}, the reflection of
waves off the boundaries inhibiting vertical motion.

To put the importance of the Poynting flux to magnetic field evolution
into context, the total Poynting flux, $\lb F_{\mathcal{P}} \rb$,
removes roughly $11\%$ of the total generated magnetic energy ($\sim
\lb \mathcal{S}_{B'} \rb + \lb \mathcal{S}_{\bar{B}} \rb$ - see
Table~\ref{tab:control_vol}), with roughly $8\%$ going into the corona
via $\lb F_{\mathcal{P}-\theta} \rb$. These results disagree with
\cite{Miller:2000}, who inferred that $\sim25\%$ of the magnetic
energy produced in the disk body is transported to the corona. There
are, however, substantial differences between the numerical setup and
methods of analysis used by \cite{Miller:2000} and ourselves, which
could account for some of this difference. Nonetheless, there is a
very important difference in the physical mechanism posited to drive
magnetic energy into the corona. \cite{Miller:2000} attribute buoyant
rising of flux tubes, whereas the results in Fig.~\ref{fig:Poynting}
and Table~\ref{tab:control_vol} suggest that it is in fact the stress
like part of the Poynting flux, not buoyant advection, that plays the
prominent role. This finding has important implications for models of
coronal powering, as discussed at the beginning of this section.

\subsection{Turbulent kinetic energy evolution}
\label{subsec:KE}

\begin{figure}
  \begin{center}
    \begin{tabular}{c}
\resizebox{80mm}{!}{\includegraphics{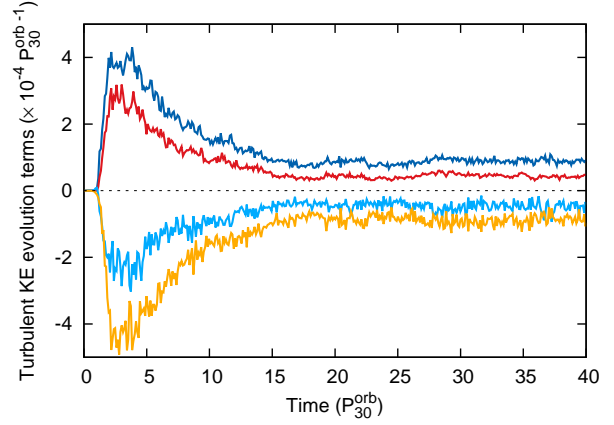}} \\
 \end{tabular}
 \caption{Terms pertaining to the evolution of the turbulent kinetic
   energy (\S~\ref{subsec:KE}). Shown are: $\lb \mathcal{S}_{K'} \rb$
   (dark blue), $- \lb \mathcal{L}' \rb$ (red), $-\llb v'_{i}
   \frac{\partial p}{\partial x_{i}} \rrb$ (light blue), and $\lb
   D^{\rm visc} \rb$ (orange). Positive values indicate energy input,
   whereas negative values indicate removal of energy.}
    \label{fig:KE_control_vol}
  \end{center}
\end{figure}

The turbulent kinetic energy provides an insight into the energy
possessed by velocity fluctuations that are effectively carried along
by the mean rotation. In this regard, when considering an analysis in
the inertial frame, it is essential to separate turbulent and mean
velocities. The derivation of the turbulent kinetic energy equation
begins with the momentum equation (Eq~\ref{eqn:mom}) with a term added
to account for numerical viscous losses, $d^{\rm visc}$,
\begin{eqnarray}
  \frac{\partial \rho v_{i}}{\partial t} + \frac{\partial}{\partial
    x_{j}}(\rho v_{i}v_{j}) = -\rho \frac{\partial \Phi}{\partial x_{i}} -
  \frac{\partial p}{\partial x_{i}} + \frac{\partial M_{ij}}{\partial
    x_{j}} + d_{i}^{\rm visc}.
\end{eqnarray}
The velocity and magnetic fields are then decomposed into mean and
turbulent components (\S~\ref{subsec:mean_turb}), the scalar product
with $v'_{i}$ is taken, and the resulting equation is Reynolds
averaged. The term containing the Maxwell stress may be expanded as
follows,
\begin{equation}
  \overline{ v'_{i} \frac{\partial M_{ij}}{\partial x_{j}} } =
  \overline{ v'_{i} \frac{\partial M'_{ij}}{\partial x_{j}} } +
  \overline{v'_{i} B'_{j} \frac{\partial \bar{B}_{i}}{\partial x_{j}} },
\end{equation}
where $M'_{ij}$ is the turbulent Maxwell stress tensor
(\S~\ref{subsec:mag_energy}). The azimuthally-averaged turbulent
kinetic energy equation then reads,
\begin{eqnarray}
  \frac{\partial \overline{u_{K'}}}{\partial t}&=& - \overline{\rho v'_{i} v'_{j} }\bar{v}_{i;j} -
  \frac{\partial}{\partial x_{j}}(\overline{u_{K'} } \bar{v}_{j}) -
  \frac{\partial}{\partial x_{j}}(\overline{u_{K'} v'_{j}}) \nonumber
  \\ &&
 + \overline{ v'_{i} \frac{\partial M'_{ij} }{\partial x_{j}} } +
 \overline{ v'_{i} B'_{j}} \bar{B}_{i;j}  - \overline{ v'_{i} \frac{\partial p}{\partial x_{i}} } +
 \overline{ v'_{i} d_{i}^{\rm visc}} \label{eqn:KE1},
\end{eqnarray}
where $u_{K'} =\frac{1}{2} \rho |v'|^2$ is the turbulent kinetic
energy. Finally, we take an average of Eq~(\ref{eqn:KE1}) over a
meridional ($r,\theta$) plane, which leads to a volume averaged
turbulent kinetic energy equation (see Eq~\ref{eqn:vol_av}):
\begin{eqnarray}
 \lb\dot{u}_{K'} \rb&=&  \lb \mathcal{S}_{K'} \rb + \lb F'_{K'} \rb + \lb F_{K'}
 \rb - \lb \mathcal{L}' \rb + \lb v'_{i} B'_{j} \bar{B}_{i;j} \rb
 \nonumber \\
 && - \llb v'_{i} \frac{\partial p}{\partial x_{i}} \rrb  + \lb D^{\rm
   visc} \rb, \label{eqn:KE2}
\end{eqnarray}
with the symbols defined as,
\begin{eqnarray}
  \lb \dot{u}_{K'}\rb &=& \frac{\partial}{\partial t}\lb u_{K'} \rb, \label{eqn:UK}\\
  \lb \mathcal{S}_{K'} \rb &=& - \lb \rho v'_{i} v'_{\rm j} \bar{v}_{i;j} \rb, \\
\lb F'_{K'} \rb &=& - \llb \frac{\partial}{\partial x_{j}} (u_{K'} v'_{j}) \rrb = - \frac{1}{V} \int_{s}
  u_{K'} v'_{j} \ts ds_{j}, \\ 
\lb F_{K'} \rb &=& - \llb \frac{\partial}{\partial x_{j}} (u_{K'} \bar{v}_{j}) \rrb = - \frac{1}{V} \int_{s}
  u_{K'} \bar{v}_{j} \ts ds_{j}, \\ 
 \lb D^{\rm visc}\rb &=& \lb v'_{i} d_{i}^{\rm visc} \rb. \label{eqn:Dvisc}
\end{eqnarray}
The energy removed by (numerical) viscous dissipation is accounted for
by the term $\lb D^{\rm visc} \rb$. Note that the value of $\lb D^{\rm
  visc} \rb$ is evaluated as the remainder required to balance
Eq~(\ref{eqn:KE2}), as all other terms may be computed explicitly. As
with the term $\lb D^{\rm res}_{B'} \rb$ in the magnetic energy
analysis, the advantage of this approach is that we can quantify
dissipation without the need to ascribe to it a specific functional
form. Time-averaged values for all of the terms in Eq~(\ref{eqn:KE2})
are noted in Table~\ref{tab:control_vol}, with the evolution of the
largest terms plotted in Fig.~\ref{fig:KE_control_vol}.

Turbulent kinetic energy input mainly comes from the interaction of
the Reynolds stress with shear in the mean velocity field ($\lb
\mathcal{S}_{K'} \rb$). Second to this is the contribution due to
turbulent Lorentz forces, $-\lb \mathcal{L}' \rb$. We recall that
$+\lb \mathcal{L}' \rb$ was responsible for a removal of turbulent
magnetic energy (see \S~\ref{subsec:mag_energy} and
Fig.~\ref{fig:ME_control_vol}), thus it embodies the majority of
turbulent kinetic energy input directly due to the magnetized
turbulence in the disk. Turning next to the energy removal, one sees
from Fig.~\ref{fig:KE_control_vol} that (numerical) viscous
dissipation is the largest sink, followed by the term $\lb v'_{i}
\frac{\partial p}{\partial x_{i}} \rb$, which represents the work done
on the turbulent fluctuations by the pressure force. In common with
the turbulent magnetic energy analysis in \S~\ref{subsec:mag_energy},
we find that advection of turbulent kinetic energy in/out of the disk
(the terms $\lb F'_{K'} \rb$ and $\lb F_{K'} \rb$) does little to the
global energy budget when compared to in-situ generation and
dissipation (see Table~\ref{tab:control_vol}).

\begin{figure}
  \begin{center}
    \begin{tabular}{c}
\resizebox{80mm}{!}{\includegraphics{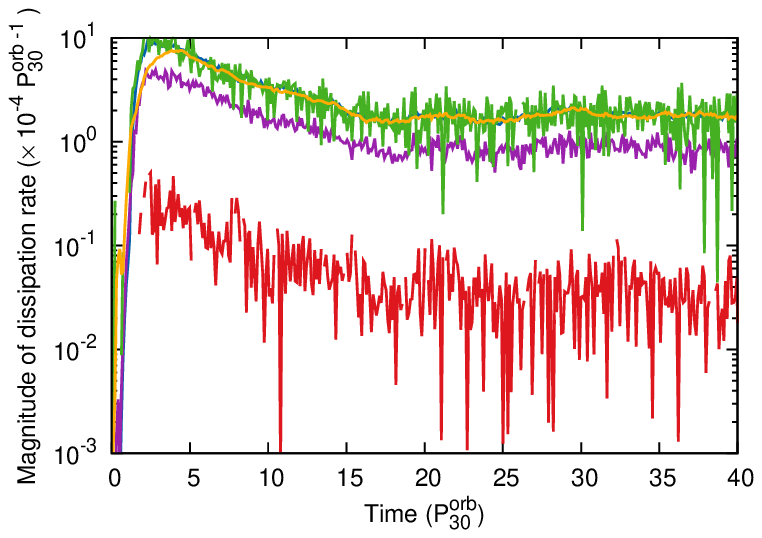}} \\
  \end{tabular}
  \caption{Comparing the various dissipation, heating, and cooling
    terms: turbulent resistive dissipation, $\lb D^{\rm res}_{B'} \rb$
    (blue); mean resistive dissipation, $\lb D^{\rm res}_{\bar{B}}
    \rb$ (red); viscous dissipation, $\lb D^{\rm visc} \rb$ (purple);
    heating, $\lb D_{\epsilon} \rb$ (green); and cooling, $\lb D_{\rm
      rad} \rb$ (orange). The curves for $\lb D^{\rm res}_{B'} \rb$,
    $\lb D_{\epsilon} \rb$, and $\lb D_{\rm rad} \rb$ are closely
    coincident on the plot. Corresponding time-averaged values can be
    found in Table~\ref{tab:control_vol}.}
    \label{fig:D}
  \end{center}
\end{figure}

\subsection{Internal energy evolution}
\label{subsec:IE}

To complete the picture of energy flow through our model disk we now
examine the internal energy equation. The volume-averaged equation is:
\begin{equation}
 \lb \dot{u}_{\epsilon} \rb = \lb F_{u_{\epsilon}} \rb + \lb F'_{u_{\epsilon}} \rb
 - \lb p v'_{k,k} \rb - \lb p \bar{v}_{k,k} \rb + \lb D^{\rm rad} \rb +
 \lb D^{\epsilon} \rb, \label{eqn:IE}
\end{equation}
where,
\begin{eqnarray}
 \lb \dot{u}_{\epsilon} \rb &=& \frac{\partial }{\partial t} \lb
 u_{\epsilon} \rb, \\
\lb F_{u_{\epsilon}} \rb &=& -\frac{1}{V} \int_s u_{\epsilon} \bar{v}_{ j} ds_{j},\\
\lb F'_{u_{\epsilon}} \rb &=& -\frac{1}{V} \int_s u_{\epsilon} v'_{j} ds_{j}.
\end{eqnarray}
Thermal energy is extracted by the cooling function, from which we
define an associated radiative loss rate:
\begin{equation}
  \lb D^{\rm rad} \rb = -\lb \rho \Lambda(T) \rb,
\end{equation}
where $\Lambda(T)$ is the cooling function which is used to drive the
temperature distribution in the disk towards that of the initial
conditions. The term $\lb D_{\epsilon} \rb$ represents numerical
heating - it takes its value from the remainder required to balance
Eq~(\ref{eqn:IE}). Time-averaged values for the terms in
Eq~(\ref{eqn:IE}) are noted in Table~\ref{tab:control_vol}.

Compressibility is important for the internal energy evolution, with
the compression-related terms $-\lb p v'_{k,k} \rb$ and $-\lb p
\bar{v}_{k,k} \rb$ having appreciable values
(Table~\ref{tab:control_vol}). A significant amount of energy is also
advected through the disk boundaries by the mean flow ($\lb
F_{u_{\epsilon}} \rb$).

Numerical heating ($\lb D^{\epsilon} \rb$) provides the largest source
of internal energy in the disk. Importantly, as we are using an ideal
MHD simulation, this shows that the energy being dissipated both
resistively and viscously by the numerics is being effectively
thermalized. Moreover, a comparison of the terms in Eq~(\ref{eqn:IE})
(see Table~\ref{tab:control_vol}), in particular $\lb D^{\epsilon}
\rb$ and $\lb D^{\rm rad} \rb$, reveals that some of the thermalized
energy is redirected into other terms, and is not entirely radiated
away. This contrasts with the standard view of accretion disks,
whereby the dissipation rates are directly equated to the radiative
emission rate \citep[e.g.][]{Shakura:1973}. The deviation is, however,
only minor, as illustrated by a comparison of dissipation, heating,
and cooling rates in Fig.~\ref{fig:D}. Of course, the description of
thermodynamics in our models is relatively simple, and a more rigorous
understanding of the flow of internal energy requires a similar
analysis to be applied to simulations using a more realistic treatment
of radiation and cooling \citep{Turner:2003, Hirose:2006, Ohsuga:2011,
  Blaes:2011, Jiang:2013, Flock:2013}. We compare our derived
dissipation rates to previous work in \S~\ref{subsec:dissipation}.

\section{Discussion}
\label{sec:discussion}

\subsection{Global energy flow}

\begin{figure}
  \begin{center}
    \begin{tabular}{c}
\resizebox{80mm}{!}{\includegraphics[angle=-90]{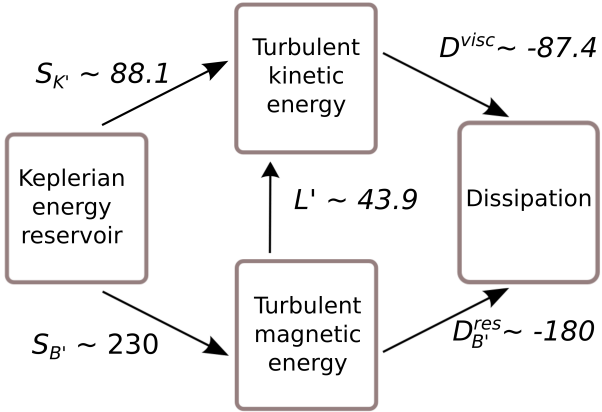}} \\
 \end{tabular}
 \caption{Energy flow diagram for turbulence in a global disk. All
   values are given in units of $10^{-6} \ts P_{30}^{\rm orb \ts -1}$
   and have been volume averaged over the disk body and time averaged
   over the interval $20<t<40 \ts P^{\rm orb}_{30}$.}
    \label{fig:energy_flow}
  \end{center}
\end{figure}

Using the values for energy production and dissipation, and the energy
exchange between turbulent magnetic and kinetic reservoirs, one can
construct a simplified energy flow diagram, which is shown in
Fig.~\ref{fig:energy_flow}. This starts with differential rotation
(shear) in the mean rotation profile, which was shown in
\S~\ref{sec:character} to be essentially Keplerian, and thus directly
linked to the gravitational potential energy. Hence, the energy source
for the largest components in the shear-stress correlations for
turbulent magnetic ($\mathcal{S}_{B'}$) and kinetic
($\mathcal{S}_{K'}$) energies can be directly connected to Keplerian
shear - see also the discussion in \cite{Brandenburg:1995},
\cite{BH98}, and \cite{Kuncic:2004}. One can write the largest
contributions to these terms as, 
\begin{equation}
  \lb \mathcal{S}_{B'} \rb \approx \llb B'_{\phi} B'_{r}
  \left(\frac{\partial v_{\rm Kep}}{\partial r} -
    \frac{v_{\rm Kep}}{r}\right )\rrb = \llb B'_{\phi} B'_{r} \frac{\d
    \Omega_{\rm Kep}}{\d \ln r}\rrb, \label{eqn:Beng}
\end{equation}
and,
\begin{equation}
  \lb \mathcal{S}_{K'} \rb \approx -\llb \rho v'_{\phi} v'_{r} \left(
    \frac{\partial v_{\rm Kep}}{\partial r} - \frac{v_{\rm Kep}}{r} \right)
  \rrb = -\llb\rho v'_{\phi} v'_{r} \frac{\d \Omega_{\rm Kep}}{\d \ln
    r}\rrb, \label{eqn:Keng}
\end{equation}
where $\Omega_{\rm Kep} = v_{\rm Kep}/r$. Combining equations (\ref{eqn:Beng})
and (\ref{eqn:Keng}) yields the well-known correlation between the
total stress tensor and shear in the the background/mean rotation
profile:
\begin{equation}
  \lb \mathcal{S}_{B'} \rb + \lb \mathcal{S}_{K'} \rb \approx -\llb \mathcal{T}_{r
    \phi} \frac{\d \Omega_{\rm Kep}}{\d \ln r} \rrb,
\end{equation}
where $\mathcal{T}_{r \phi} = \rho v'_{r} v'_{\phi} -
B'_{r} B'_{\phi}$ is the $r-\phi$ component of the total
turbulent stress tensor. As discussed by \cite{BH98}, the energy input
contributed by the above, in combination with effective angular
momentum transport, is at the heart of self-sustaining MHD turbulence
in accretion disks.

Subsequent energy exchange between turbulent magnetic and kinetic
energies largely occurs as a result of turbulent Lorentz forces,
$\mathcal{L}'$. Hence, the turbulent kinetic energy receives $\sim
1/3$ of its power from $\mathcal{L}'$. Energy flow ends with
dissipation, and we can account for resistive and viscous dissipation
with $\lb D^{\rm res}_{B'} \rb$ and $\lb D^{\rm visc} \rb$,
respectively.

There are a number of similarities, and some important differences
between Fig.~\ref{fig:energy_flow} and the energy flow chart presented
by \cite{Brandenburg:1995}. Firstly, we agree that Keplerian shear is
the main source of energy, consistent with the expectation for MRI
driven turbulence \citep{BH92}. Also, we find agreement that
essentially all energy extracted from Keplerian shear is dissipated
within the disk. However, our results indicate that turbulent kinetic
energy extracts most of its power from Keplerian rotation, in contrast
to \cite{Brandenburg:1995} who inferred that the majority of power
came from turbulent magnetic energy, with only $\sim 1/3$ being drawn
from Keplerian shear\footnote{\cite{Simon:2009} have presented an
  analysis of energetics based on unstratified shearing-box
  models. However, due to uncertainties in their values at low
  wavenumbers (i.e. on the large spatial scales that are best
  described by the volume-averaged approach in this work) it is
  difficult to make more than a speculative comparison against their
  results. Nevertheless, the transfer function analysis presented by
  \cite{Simon:2009} does suggests that Reynolds stresses are effective
  at injecting turbulent kinetic energy on large scales.}. The origin
of this difference is not clear as a comparison of energy generation
terms indicates similar results. For instance, in the analysis by
\cite{Brandenburg:1995}, $\mathcal{S}_{K'} \sim 3/2 \Omega_{0} \lb
\rho v_{x} v_{y} \rb$ and $\mathcal{S}_{B'} \sim -3/2 \Omega_{0}\lb
B_{x} B_{y} \rb$ (in cgs units, and where $x$ and $y$ are the radial
and azimuthal shearing-box coordinates and $\Omega_{0}$ is the angular
velocity of the box centre), such that the ratio of energy injection
terms is equivalent to the Maxwell-to-Reynolds stress ratio,
$\mathcal{S}_{B'}/\mathcal{S}_{K'} \equiv -\lb B_{x}B_{y} \rb / \lb
\rho v_{x} v_{y} \rb$. For model ``A'' from \cite{Brandenburg:1995} -
which includes a cooling term and is therefore a suitable candidate
for comparison - a value of $-\lb B_{x}B_{y} \rb / \lb \rho v_{x}
v_{y} \rb \simeq 3.6 $ is noted. For the global disk analysis
presented in this work, $\mathcal{S}_{B'}/\mathcal{S}_{K'} = 2.61$ and
$-\lb B'_{r}B'_{\phi} \rb / \lb \rho v'_{r} v'_{\phi} \rb = 2.69$ (see
Tables~\ref{tab:global_models} and \ref{tab:control_vol}). (The minor
difference between $\mathcal{S}_{B'}/\mathcal{S}_{K'}$ and $-\lb
B'_{r}B'_{\phi} \rb / \lb \rho v'_{r} v'_{\phi} \rb $ due to
components in the energy injection terms in addition to the $r-\phi$
Reynolds/Maxwell stress.) Hence, the reasons for the disagreement in
energy flow diagrams is not immediately apparent. A possible
resolution would be to repeat the mean-field analysis presented in
this work for a well-resolved stratified shearing-box. This is beyond
the scope of the present work, but would be a useful subject for
future studies.

In summary, $81\%$ of the energy extracted from Keplerian shear in the
disk body is dissipated; this is estimated from $(\lb D^{\rm res}_{B'}
\rb + \lb D^{\rm res}_{\bar{B}} \rb + \lb D^{\rm visc} \rb )/(\lb
\mathcal{S}_{B'} \rb + \lb \mathcal{S}_{\bar{B}} \rb +\lb
\mathcal{S}_{K'} \rb)$. This number is considerable, and indicates
that the disk is very efficient at thermalizing kinetic and magnetic
energy.

\subsection{Comparison with previous studies}
\label{subsec:dissipation}

The vast majority of previous studies of energetics in a turbulent
magnetized disk have utilised local (shearing-box) simulations. Here
we compare and contrast those results against our findings from a
global disk model. Our results agree well with those of
\cite{Simon:2009}, who reported a viscous-to-resistive dissipation
ratio, $\lb D^{\rm visc} \rb / \lb D^{\rm res} \rb \simeq 0.5$ for
unstratified shearing-box models with purely numerical dissipation,
which should be compared with $\lb D^{\rm visc} \rb / (\lb D^{\rm
  res}_{B'} \rb + \lb D^{\rm res}_{\bar{B}} \rb)\simeq 0.48$ from this
work. Making comparisons against dissipation rates for models which
have included explicit diffusion coefficients is somewhat difficult
and confusing. \cite{Miller:2000} found that resistive dissipation was
$0.71\%$ of viscous dissipation in their stratified disk models. In
contrast, \cite{Brandenburg:1995} note $\lb D^{\rm visc} \rb/ \lb
D^{\rm res} \rb \simeq 4/3$. One possible explanation for the
difference between these two models would be that they have different
effective magnetic Prandtl numbers, $Pr_{\rm M} = \nu/ \eta$, where
$\nu$ and $\eta$ are viscosity and resistivity, respectively. In this
regard, the models of \cite{Brandenburg:1995} have $Pr_{\rm M}\sim1$
if based on explicit diffusion coefficients, and $Pr_{\rm M}\sim10$ if
based on artificial viscosity coefficients. However, viscosity and
resistivity values used by \cite{Miller:2000} are not given,
complicating a comparison. In these somewhat earlier simulations of
MRI-driven turbulence, the focus was less on the influence of the
Prandtl number dependence of the turbulent dissipation, and more on
exploring fundamental features of the turbulence. In recent years the
focus has shifted somewhat, with a Prandtl number dependence to the
turbulence having been revealed \citep{Fromang:2007b, Lesur:2007,
  Longaretti:2010, Simon:2011, Kapyla:2011, Oishi:2011}. Examining the
role of the Prandtl number in setting the dissipation rates in a
global context is, therefore, an interesting avenue for future
studies.

We have not defined a functional form for the numerical dissipation
terms in this work. Indeed, the adopted analysis approach does not
demand it. Numerical dissipation does not necessarily behave in the
same manner as standard non-ideal dissipation. This point is discussed
further by \cite{Xu:2001}, \cite{Hirose:2006}, \cite{Fromang:2007a},
\cite{Simon:2009}, \cite{Hawley:2011}, \cite{Parkin:2013b}, and
\cite{Salvesen:2013}. Nevertheless, if we assume an Ohmic form for
$\lb D^{\rm res}_{B'} \rb$, and a Newtonian shear-stress form for $\lb
D^{\rm visc} \rb$, we find a numerical magnetic Prandtl number,
$Pr_{\rm M} \simeq 0.6$. Although this does not agree exactly with the
value of $Pr_{\rm M} \simeq 2$ estimated by \cite{Simon:2009}, both
values are within a factor of two of unity. Considering that
astrophysical objects are expected to span a wide range of values of
$Pr_{\rm M}$ \citep{Brandenburg:2005b, Balbus:2008}, deviating
substantially from unity in many cases, further progress requires the
consideration of explicitly non-ideal global models
\citep[e.g.][]{Flock:2012b}.

\section{Conclusions}
\label{sec:conclusions}

The results of a high-resolution, three-dimensional,
magnetohydrodynamic disk simulation have been used to evaluate the
flow of energy, and detailed interactions between mean and turbulent
fields, for MRI-driven turbulence in a global setting. To this end we
have used a Reynolds averaged mean-field approach to study equations
for: the turbulent magnetic energy, mean magnetic energy, mean field
induction, turbulent kinetic energy, and internal energy. The results
show that the correlation between turbulent Reynolds and Maxwell
stresses with shear in the (almost Keplerian) mean rotation provides
the main source of turbulent energy. Turbulent Lorentz forces extract
magnetic energy and inject it into kinetic energy. Terms featuring the
mean magnetic fields do not contribute significantly to the turbulent
magnetic energy evolution. However, turbulent fields are fluctuations
about a mean. As such, the indirect influence of mean magnetic fields
on the turbulent magnetic energy can be understood from the mean
magnetic field induction equation.  For the case of the mean radial
magnetic field the turbulent EMF is the primary driver, highlighting
an example of turbulence driving the evolution of mean fields.

A number of interesting results have also been revealed regarding the
Poynting flux. During the quasi-steady state of the simulation, the
Poynting flux carries an amount of energy equating to $8\%$ of the
total magnetic energy input into the corona. This value is somewhat
lower than estimates from previous shearing-box studies, although this
could simply be the result of the different simulation setup and
analysis methods used. More important, however, is the finding that
the stress-related part of the Poynting flux is roughly seven times
larger than than the advection-related part, a result which has
considerable importance for models of coronal powering in Seyfert
galaxies that typically invoke buoyant rising of flux tubes for
electromagnetic energy transport.

The basic picture for the disk ``engine'' painted by the results of
the analysis in this work bears many similarities with findings from
previous shearing-box studies: stress-shear terms power the turbulence
with turbulent Lorentz forces providing a means of transporting energy
between the magnetic and kinetic reservoirs, and all paths ending with
dissipation. However, a number of important differences have been
uncovered. Most striking is our finding that the turbulent kinetic
energy is mostly fed from terms associated with Keplerian shear, and
not by Lorentz forces, in contrast to the results of
\cite{Brandenburg:1995}.

In closing we note that the control volume analysis utilised in this
work provides spatially-averaged information about turbulent
energetics. However, the power spectra presented in
Fig.~\ref{fig:Ek_dB_dv} indicate that the ordering of strengths in
directional components of the turbulent fields changes between the
very largest and smallest scales. Hence, with an eye to gaining
further insight into the scale-dependence of energetics in a turbulent
disk, and the possibility that the very smallest scales behave
differently to the largest (energy containing) scales, it would be
interesting to perform a follow-up analysis in wavenumber space.

\subsection*{Acknowledgements}
I thank Geoffrey Bicknell for numerous insightful discussions and the
anonymous refereee for a prompt and constructive report which helped
to clarify points in the paper. This research was supported under the
Australian Research Council's Discovery Projects funding scheme
(project number DP1096417), and by NCI Facility at the ANU.


\begin{thebibliography}{}

\bibitem[\protect\citeauthoryear{{Arlt} \& {R{\"u}diger}}{{Arlt} \&
  {R{\"u}diger}}{2001}]{Arlt:2001}
{Arlt}, R. \& {R{\"u}diger}, G. 2001, \aap, 374, 1035

\bibitem[\protect\citeauthoryear{{Bai} \& {Stone}}{{Bai} \&
  {Stone}}{2013}]{Bai:2013}
{Bai}, X.-N. \& {Stone}, J.~M. 2013, \apj, 767, 30

\bibitem[\protect\citeauthoryear{{Balbus} \& {Hawley}}{{Balbus} \&
  {Hawley}}{1991}]{BH91}
{Balbus}, S.~A. \& {Hawley}, J.~F. 1991, \apj, 376, 214

\bibitem[\protect\citeauthoryear{{Balbus} \& {Hawley}}{{Balbus} \&
  {Hawley}}{1992}]{BH92}
{Balbus}, S.~A. \& {Hawley}, J.~F. 1992, \apj, 400, 610

\bibitem[\protect\citeauthoryear{{Balbus} \& {Hawley}}{{Balbus} \&
  {Hawley}}{1998}]{BH98}
{Balbus}, S.~A. \& {Hawley}, J.~F. 1998, Reviews of Modern Physics, 70, 1

\bibitem[\protect\citeauthoryear{{Balbus} \& {Henri}}{{Balbus} \&
  {Henri}}{2008}]{Balbus:2008}
{Balbus}, S.~A. \& {Henri}, P. 2008, \apj, 674, 408

\bibitem[\protect\citeauthoryear{{Beckwith}, {Armitage} \& {Simon}}{{Beckwith}
  et~al.}{2011}]{Beckwith:2011}
{Beckwith}, K., {Armitage}, P.~J., \& {Simon}, J.~B. 2011, \mnras, 416, 361

\bibitem[\protect\citeauthoryear{{Blackman} \& {Pessah}}{{Blackman} \&
  {Pessah}}{2009}]{Blackman:2009}
{Blackman}, E.~G. \& {Pessah}, M.~E. 2009, \apjl, 704, L113

\bibitem[\protect\citeauthoryear{{Blaes}, {Krolik}, {Hirose} \&
  {Shabaltas}}{{Blaes} et~al.}{2011}]{Blaes:2011}
{Blaes}, O., {Krolik}, J.~H., {Hirose}, S., \& {Shabaltas}, N. 2011, \apj, 733,
  110

\bibitem[\protect\citeauthoryear{{Blandford} \& {Payne}}{{Blandford} \&
  {Payne}}{1982}]{Blandford:1982}
{Blandford}, R.~D. \& {Payne}, D.~G. 1982, \mnras, 199, 883

\bibitem[\protect\citeauthoryear{{Brandenburg}}{{Brandenburg}}{2005}]{Brandenburg:2005}
{Brandenburg}, A. 2005, Astronomische Nachrichten, 326, 787

\bibitem[\protect\citeauthoryear{{Brandenburg}, {Nordlund}, {Stein} \&
  {Torkelsson}}{{Brandenburg} et~al.}{1995}]{Brandenburg:1995}
{Brandenburg}, A., {Nordlund}, A., {Stein}, R.~F., \& {Torkelsson}, U. 1995,
  \apj, 446, 741

\bibitem[\protect\citeauthoryear{{Brandenburg} \& {Subramanian}}{{Brandenburg}
  \& {Subramanian}}{2005}]{Brandenburg:2005b}
{Brandenburg}, A. \& {Subramanian}, K. 2005, \physrep, 417, 1

\bibitem[\protect\citeauthoryear{{Colella} \& {Woodward}}{{Colella} \&
  {Woodward}}{1984}]{Colella:1984}
{Colella}, P. \& {Woodward}, P.~R. 1984, \jcp, 54, 174

\bibitem[\protect\citeauthoryear{Davidson}{Davidson}{2004}]{Davidson:2004}
Davidson, P. 2004, Turbulence : An Introduction for Scientists and Engineers:
  An Introduction for Scientists and Engineers.
OUP Oxford

\bibitem[\protect\citeauthoryear{{Davis}, {Stone} \& {Pessah}}{{Davis}
  et~al.}{2010}]{Davis:2010}
{Davis}, S.~W., {Stone}, J.~M., \& {Pessah}, M.~E. 2010, \apj, 713, 52

\bibitem[\protect\citeauthoryear{Favre}{Favre}{1969}]{Favre:1969}
Favre, A. 1969, in Problems of hydrodynamics and continuum mechanics.
SIAM

\bibitem[\protect\citeauthoryear{{Field} \& {Rogers}}{{Field} \&
  {Rogers}}{1993}]{Field:1993}
{Field}, G.~B. \& {Rogers}, R.~D. 1993, \apj, 403, 94

\bibitem[\protect\citeauthoryear{{Flock}, {Dzyurkevich}, {Klahr}, {Turner} \&
  {Henning}}{{Flock} et~al.}{2012a}]{Flock:2012}
{Flock}, M., {Dzyurkevich}, N., {Klahr}, H., {Turner}, N., \& {Henning}, T.
  2012a, \apj, 744, 144

\bibitem[\protect\citeauthoryear{{Flock}, {Dzyurkevich}, {Klahr}, {Turner} \&
  {Henning}}{{Flock} et~al.}{2011}]{Flock:2011}
{Flock}, M., {Dzyurkevich}, N., {Klahr}, H., {Turner}, N.~J., \& {Henning}, T.
  2011, \apj, 735, 122

\bibitem[\protect\citeauthoryear{{Flock}, {Fromang}, {Gonz{\'a}lez} \&
  {Commer{\c c}on}}{{Flock} et~al.}{2013}]{Flock:2013}
{Flock}, M., {Fromang}, S., {Gonz{\'a}lez}, M., \& {Commer{\c c}on}, B. 2013,
  arXiv:1310.5865

\bibitem[\protect\citeauthoryear{{Flock}, {Henning} \& {Klahr}}{{Flock}
  et~al.}{2012b}]{Flock:2012b}
{Flock}, M., {Henning}, T., \& {Klahr}, H. 2012b, \apj, 761, 95

\bibitem[\protect\citeauthoryear{{Fromang}, {Latter}, {Lesur} \&
  {Ogilvie}}{{Fromang} et~al.}{2013}]{Fromang:2013}
{Fromang}, S., {Latter}, H., {Lesur}, G., \& {Ogilvie}, G.~I. 2013, \aap, 552,
  A71

\bibitem[\protect\citeauthoryear{{Fromang} \& {Nelson}}{{Fromang} \&
  {Nelson}}{2006}]{Fromang:2006}
{Fromang}, S. \& {Nelson}, R.~P. 2006, \aap, 457, 343

\bibitem[\protect\citeauthoryear{{Fromang} \& {Papaloizou}}{{Fromang} \&
  {Papaloizou}}{2007}]{Fromang:2007a}
{Fromang}, S. \& {Papaloizou}, J. 2007, \aap, 476, 1113

\bibitem[\protect\citeauthoryear{{Fromang}, {Papaloizou}, {Lesur} \&
  {Heinemann}}{{Fromang} et~al.}{2007}]{Fromang:2007b}
{Fromang}, S., {Papaloizou}, J., {Lesur}, G., \& {Heinemann}, T. 2007, \aap,
  476, 1123

\bibitem[\protect\citeauthoryear{{Gardiner} \& {Stone}}{{Gardiner} \&
  {Stone}}{2005a}]{Gardiner:2005}
{Gardiner}, T.~A. \& {Stone}, J.~M. 2005a, Journal of Computational Physics,
  205, 509

\bibitem[\protect\citeauthoryear{{Gardiner} \& {Stone}}{{Gardiner} \&
  {Stone}}{2005b}]{Gardiner:2005b}
{Gardiner}, T.~A. \& {Stone}, J.~M. 2005b, in {de Gouveia dal Pino} E.~M.,
  {Lugones} G.,   {Lazarian} A.,  eds, Magnetic Fields in the Universe: From
  Laboratory and Stars to Primordial Structures. Vol.~784 of AIP
  Conf. Ser. pp 475--488

\bibitem[\protect\citeauthoryear{{Gombosi}, {T{\'o}th}, {de Zeeuw}, {Hansen},
  {Kabin} \& {Powell}}{{Gombosi} et~al.}{2002}]{Gombosi:2002}
{Gombosi}, T.~I., {T{\'o}th}, G., {de Zeeuw}, D.~L., {Hansen}, K.~C., {Kabin},
  K., \& {Powell}, K.~G. 2002, Journal of Computational Physics, 177, 176

\bibitem[\protect\citeauthoryear{{Gressel}}{{Gressel}}{2010}]{Gressel:2010}
{Gressel}, O. 2010, \mnras, 405, 41

\bibitem[\protect\citeauthoryear{{Guan} \& {Gammie}}{{Guan} \&
  {Gammie}}{2011}]{Guan:2011}
{Guan}, X. \& {Gammie}, C.~F. 2011, \apj, 728, 130

\bibitem[\protect\citeauthoryear{{Haardt} \& {Maraschi}}{{Haardt} \&
  {Maraschi}}{1991}]{Haardt:1991}
{Haardt}, F. \& {Maraschi}, L. 1991, \apjl, 380, L51

\bibitem[\protect\citeauthoryear{{Haardt} \& {Maraschi}}{{Haardt} \&
  {Maraschi}}{1993}]{Haardt:1993}
{Haardt}, F. \& {Maraschi}, L. 1993, \apj, 413, 507

\bibitem[\protect\citeauthoryear{{Hawley}, {Gammie} \& {Balbus}}{{Hawley}
  et~al.}{1995}]{Hawley:1995}
{Hawley}, J.~F., {Gammie}, C.~F., \& {Balbus}, S.~A. 1995, \apj, 440, 742

\bibitem[\protect\citeauthoryear{{Hawley}, {Guan} \& {Krolik}}{{Hawley}
  et~al.}{2011}]{Hawley:2011}
{Hawley}, J.~F., {Guan}, X., \& {Krolik}, J.~H. 2011, \apj, 738, 84

\bibitem[\protect\citeauthoryear{{Hawley}, {Richers}, {Guan} \&
  {Krolik}}{{Hawley} et~al.}{2013}]{Hawley:2013}
{Hawley}, J.~F., {Richers}, S.~A., {Guan}, X., \& {Krolik}, J.~H. 2013, \apj,
  772, 102

\bibitem[\protect\citeauthoryear{{Heinemann} \& {Papaloizou}}{{Heinemann} \&
  {Papaloizou}}{2009}]{Heinemann_Papaloizou:2009}
{Heinemann}, T. \& {Papaloizou}, J.~C.~B. 2009, \mnras, 397, 64

\bibitem[\protect\citeauthoryear{{Heinemann} \& {Papaloizou}}{{Heinemann} \&
  {Papaloizou}}{2012}]{Heinemann_Papaloizou:2012}
{Heinemann}, T. \& {Papaloizou}, J.~C.~B. 2012, \mnras, 419, 1085

\bibitem[\protect\citeauthoryear{{Hirose}, {Krolik} \& {Stone}}{{Hirose}
  et~al.}{2006}]{Hirose:2006}
{Hirose}, S., {Krolik}, J.~H., \& {Stone}, J.~M. 2006, \apj, 640, 901

\bibitem[\protect\citeauthoryear{{Io} \& {Suzuki}}{{Io} \&
  {Suzuki}}{2013}]{Io:2013}
{Io}, Y. \& {Suzuki}, T.~K. 2013, arXiv:1308.6427

\bibitem[\protect\citeauthoryear{{Jiang}, {Stone} \& {Davis}}{{Jiang}
  et~al.}{2013}]{Jiang:2013}
{Jiang}, Y.-F., {Stone}, J.~M., \& {Davis}, S.~W. 2013, \apj, 778, 65

\bibitem[\protect\citeauthoryear{{Johansen}, {Youdin} \& {Klahr}}{{Johansen}
  et~al.}{2009}]{Johansen:2009}
{Johansen}, A., {Youdin}, A., \& {Klahr}, H. 2009, \apj, 697, 1269

\bibitem[\protect\citeauthoryear{{K{\"a}pyl{\"a}} \& {Korpi}}{{K{\"a}pyl{\"a}}
  \& {Korpi}}{2011}]{Kapyla:2011}
{K{\"a}pyl{\"a}}, P.~J. \& {Korpi}, M.~J. 2011, \mnras, 413, 901

\bibitem[\protect\citeauthoryear{{Krause} \& {Raedler}}{{Krause} \&
  {Raedler}}{1980}]{Krause:1980}
{Krause}, F. \& {Raedler}, K.-H. 1980, {Mean-field magnetohydrodynamics and
  dynamo theory}

\bibitem[\protect\citeauthoryear{{Kuncic} \& {Bicknell}}{{Kuncic} \&
  {Bicknell}}{2004}]{Kuncic:2004}
{Kuncic}, Z. \& {Bicknell}, G.~V. 2004, \apj, 616, 669

\bibitem[\protect\citeauthoryear{{Lesur}, {Ferreira} \& {Ogilvie}}{{Lesur}
  et~al.}{2013}]{Lesur:2013}
{Lesur}, G., {Ferreira}, J., \& {Ogilvie}, G.~I. 2013, \aap, 550, A61

\bibitem[\protect\citeauthoryear{{Lesur} \& {Longaretti}}{{Lesur} \&
  {Longaretti}}{2007}]{Lesur:2007}
{Lesur}, G. \& {Longaretti}, P.-Y. 2007, \mnras, 378, 1471

\bibitem[\protect\citeauthoryear{{Lesur} \& {Ogilvie}}{{Lesur} \&
  {Ogilvie}}{2008}]{Lesur_Ogilvie:2008b}
{Lesur}, G. \& {Ogilvie}, G.~I. 2008, \mnras, 391, 1437

\bibitem[\protect\citeauthoryear{{Longaretti} \& {Lesur}}{{Longaretti} \&
  {Lesur}}{2010}]{Longaretti:2010}
{Longaretti}, P.-Y. \& {Lesur}, G. 2010, \aap, 516, A51

\bibitem[\protect\citeauthoryear{{Mignone}, {Bodo}, {Massaglia}, {Matsakos},
  {Tesileanu}, {Zanni} \& {Ferrari}}{{Mignone} et~al.}{2007}]{Mignone:2007}
{Mignone}, A., {Bodo}, G., {Massaglia}, S., {Matsakos}, T., {Tesileanu}, O.,
  {Zanni}, C., \& {Ferrari}, A. 2007, \apjs, 170, 228

\bibitem[\protect\citeauthoryear{{Mignone}, {Flock}, {Stute}, {Kolb} \&
  {Muscianisi}}{{Mignone} et~al.}{2012}]{Mignone:2012}
{Mignone}, A., {Flock}, M., {Stute}, M., {Kolb}, S.~M., \& {Muscianisi}, G.
  2012, \aap, 545, A152

\bibitem[\protect\citeauthoryear{{Miller} \& {Stone}}{{Miller} \&
  {Stone}}{2000}]{Miller:2000}
{Miller}, K.~A. \& {Stone}, J.~M. 2000, \apj, 534, 398

\bibitem[\protect\citeauthoryear{{Miyoshi} \& {Kusano}}{{Miyoshi} \&
  {Kusano}}{2005}]{Miyoshi:2005}
{Miyoshi}, T. \& {Kusano}, K. 2005, \jcp, 208, 315

\bibitem[\protect\citeauthoryear{{Noble}, {Krolik} \& {Hawley}}{{Noble}
  et~al.}{2010}]{Noble:2010}
{Noble}, S.~C., {Krolik}, J.~H., \& {Hawley}, J.~F. 2010, \apj, 711, 959

\bibitem[\protect\citeauthoryear{{Ogilvie}}{{Ogilvie}}{2003}]{Ogilvie:2003}
{Ogilvie}, G.~I. 2003, \mnras, 340, 969

\bibitem[\protect\citeauthoryear{{Ogilvie}}{{Ogilvie}}{2012}]{Ogilvie:2012}
{Ogilvie}, G.~I. 2012, \mnras, 423, 1318

\bibitem[\protect\citeauthoryear{{Ohsuga} \& {Mineshige}}{{Ohsuga} \&
  {Mineshige}}{2011}]{Ohsuga:2011}
{Ohsuga}, K. \& {Mineshige}, S. 2011, \apj, 736, 2

\bibitem[\protect\citeauthoryear{{Oishi} \& {Mac Low}}{{Oishi} \& {Mac
  Low}}{2011}]{Oishi:2011}
{Oishi}, J.~S. \& {Mac Low}, M.-M. 2011, \apj, 740, 18

\bibitem[\protect\citeauthoryear{{O'Neill}, {Reynolds}, {Miller} \&
  {Sorathia}}{{O'Neill} et~al.}{2011}]{O'Neill:2011}
{O'Neill}, S.~M., {Reynolds}, C.~S., {Miller}, M.~C., \& {Sorathia}, K.~A.
  2011, \apj, 736, 107

\bibitem[\protect\citeauthoryear{{Paczy{\'n}sky} \& {Wiita}}{{Paczy{\'n}sky} \&
  {Wiita}}{1980}]{Paczynsky:1980}
{Paczy{\'n}sky}, B. \& {Wiita}, P.~J. 1980, \aap, 88, 23

\bibitem[\protect\citeauthoryear{{Parkin} \& {Bicknell}}{{Parkin} \&
  {Bicknell}}{2013a}]{Parkin:2013}
{Parkin}, E.~R. \& {Bicknell}, G.~V. 2013a, \apj, 763, 99

\bibitem[\protect\citeauthoryear{{Parkin} \& {Bicknell}}{{Parkin} \&
  {Bicknell}}{2013b}]{Parkin:2013b}
{Parkin}, E.~R. \& {Bicknell}, G.~V. 2013b, \mnras, 435, 2281

\bibitem[\protect\citeauthoryear{{Pessah}, {Chan} \& {Psaltis}}{{Pessah}
  et~al.}{2006}]{Pessah:2006}
{Pessah}, M.~E., {Chan}, C.-K., \& {Psaltis}, D. 2006, \mnras, 372, 183

\bibitem[\protect\citeauthoryear{{Rider}, {Greenough} \& {Kamm}}{{Rider}
  et~al.}{2007}]{Rider:2007}
{Rider}, W.~J., {Greenough}, J.~A., \& {Kamm}, J.~R. 2007, \jcp, 225, 1827

\bibitem[\protect\citeauthoryear{{Salvesen}, {Beckwith}, {Simon}, {O'Neill} \&
  {Begelman}}{{Salvesen} et~al.}{2013}]{Salvesen:2013}
{Salvesen}, G., {Beckwith}, K., {Simon}, J.~B., {O'Neill}, S.~M., \&
  {Begelman}, M.~C. 2013, arXiv:1303.5052

\bibitem[\protect\citeauthoryear{{Shakura} \& {Sunyaev}}{{Shakura} \&
  {Sunyaev}}{1973}]{Shakura:1973}
{Shakura}, N.~I. \& {Sunyaev}, R.~A. 1973, \aap, 24, 337

\bibitem[\protect\citeauthoryear{{Shi}, {Krolik} \& {Hirose}}{{Shi}
  et~al.}{2010}]{Shi:2010}
{Shi}, J., {Krolik}, J.~H., \& {Hirose}, S. 2010, \apj, 708, 1716

\bibitem[\protect\citeauthoryear{{Simon}, {Hawley} \& {Beckwith}}{{Simon}
  et~al.}{2009}]{Simon:2009}
{Simon}, J.~B., {Hawley}, J.~F., \& {Beckwith}, K. 2009, \apj, 690, 974

\bibitem[\protect\citeauthoryear{{Simon}, {Hawley} \& {Beckwith}}{{Simon}
  et~al.}{2011}]{Simon:2011}
{Simon}, J.~B., {Hawley}, J.~F., \& {Beckwith}, K. 2011, \apj, 730, 94

\bibitem[\protect\citeauthoryear{{Sorathia}, {Reynolds} \&
  {Armitage}}{{Sorathia} et~al.}{2010}]{Sorathia:2010}
{Sorathia}, K.~A., {Reynolds}, C.~S., \& {Armitage}, P.~J. 2010, \apj, 712,
  1241

\bibitem[\protect\citeauthoryear{{Sorathia}, {Reynolds}, {Stone} \&
  {Beckwith}}{{Sorathia} et~al.}{2012}]{Sorathia:2012}
{Sorathia}, K.~A., {Reynolds}, C.~S., {Stone}, J.~M., \& {Beckwith}, K. 2012,
  \apj, 749, 189

\bibitem[\protect\citeauthoryear{{Stone}, {Hawley}, {Gammie} \&
  {Balbus}}{{Stone} et~al.}{1996}]{Stone:1996}
{Stone}, J.~M., {Hawley}, J.~F., {Gammie}, C.~F., \& {Balbus}, S.~A. 1996,
  \apj, 463, 656

\bibitem[\protect\citeauthoryear{{Suzuki} \& {Inutsuka}}{{Suzuki} \&
  {Inutsuka}}{2009}]{Suzuki:2009}
{Suzuki}, T.~K. \& {Inutsuka}, S.-i. 2009, \apjl, 691, L49

\bibitem[\protect\citeauthoryear{{Suzuki} \& {Inutsuka}}{{Suzuki} \&
  {Inutsuka}}{2013}]{Suzuki:2013}
{Suzuki}, T.~K. \& {Inutsuka}, S.-i. 2013, arXiv:1309.6916

\bibitem[\protect\citeauthoryear{{Suzuki}, {Muto} \& {Inutsuka}}{{Suzuki}
  et~al.}{2010}]{Suzuki:2010}
{Suzuki}, T.~K., {Muto}, T., \& {Inutsuka}, S.-i. 2010, \apj, 718, 1289

\bibitem[\protect\citeauthoryear{{Terquem} \& {Papaloizou}}{{Terquem} \&
  {Papaloizou}}{1996}]{Terquem:1996}
{Terquem}, C. \& {Papaloizou}, J.~C.~B. 1996, \mnras, 279, 767

\bibitem[\protect\citeauthoryear{{Tout} \& {Pringle}}{{Tout} \&
  {Pringle}}{1992}]{Tout:1992}
{Tout}, C.~A. \& {Pringle}, J.~E. 1992, \mnras, 259, 604

\bibitem[\protect\citeauthoryear{{Turner}, {Stone}, {Krolik} \&
  {Sano}}{{Turner} et~al.}{2003}]{Turner:2003}
{Turner}, N.~J., {Stone}, J.~M., {Krolik}, J.~H., \& {Sano}, T. 2003, \apj,
  593, 992

\bibitem[\protect\citeauthoryear{{Uzdensky}}{{Uzdensky}}{2013}]{Uzdensky:2013}
{Uzdensky}, D.~A. 2013, \apj, 775, 103

\bibitem[\protect\citeauthoryear{{Xu} \& {Li}}{{Xu} \& {Li}}{2001}]{Xu:2001}
{Xu}, K. \& {Li}, Z. 2001, International Journal for Numerical Methods in
  Fluids, 37, 1

\end{thebibliography}

 \newcommand{\noop}[1]{}

\label{lastpage}

\end{document}